\documentclass[aps,prb,preprint,groupedaddress,showpacs]{revtex4}
\usepackage[dvips]{graphicx}
\newcommand{\be}{\begin{equation}}
\newcommand{\ee}{\end{equation}}
\newcommand{\bea}{\begin{eqnarray}}
\newcommand{\eea}{\end{eqnarray}}

\begin{document}
\title{Absence of self-averaging in the complex admittance for 
transport through disordered media
}

\author{Mitsuhiro Kawasaki\cite{byline}}

\author{Takashi Odagaki}
\affiliation{Department of Physics, Kyushu University, Fukuoka 812-8581, Japan
}

\author{Klaus W. Kehr\cite{byline2}}
\affiliation{
Institute f\"{u}r Festk\"{o}rperforschung, Forschungszentrum J\"{u}lich GmbH, 
D-42525 J\"{u}lich, Germany
}
\date{\today}
\begin{abstract}
Random walk models in one-dimensional disordered media 
with an oscillatory input current are investigated theoretically as 
generic models of the boundary perturbation experiment. 
It is shown that the complex admittance obtained in the experiment 
is not self-averaging when the jump rates $w_i$ are random variables
with the power-law distribution $\rho(w_i)\sim {w_i}^{\alpha-1} 
\ (0 < \alpha \leq 1)$. 
More precisely, the frequency-dependence of the disorder-averaged 
admittance $\langle \chi \rangle$ disagrees with that of the admittance
$\chi$ of any sample. It implies that the Cole-Cole plot of $\langle
\chi \rangle$ shows a different shape from that of the Cole-Cole plots
of $\chi$ of each sample. The condition for absence of self-averaging is
investigated with a toy model in terms of the extended central limit 
theorem. Higher dimensional media are also investigated and it is shown 
that the complex admittance for two-dimensional or three-dimensional 
media is also non-self-averaging.
\end{abstract}

\maketitle

\section{INTRODUCTION}
\label{sec:introduction}

Disordered systems like amorphous semiconductors have a large number of 
remarkable properties inherent in their disorder, like the dispersive 
transport \cite{Scher91}. Hence, the disordered systems 
have wide areas of application like the photoreceptors of 
solar batteries and photocopiers \cite{Scharfe70}. 
So, they attract continuous research interests. 
In investigation of such remarkable properties, 
two kinds of statistics are to be considered. 
One is of course statistics of thermal fluctuation, which is considered 
in traditional statistical physics. 
The other is statistics of random structure of disordered systems. 
In order to treat this type of statistics, 
random structure is described with random variables and the probability
distribution function of the random variables characterizes the substances.

It is usually assumed in studies of disordered systems 
that a sample used in experiments is sufficiently large. 
Hence, a measurement of any observable in such
a system corresponds to an average over the ensemble of all realizations
of the disorder. Quantities for which this assumption is valid are said to be 
self-averaging. The assumption means no sample dependence in the quantities. 
So, if the assumption is valid, reproducibility of experimental results for 
different samples is guaranteed. 
In addition, if the assumption is valid, 
in theoretical analyses only a disorder-averaged
quantity is to be calculated since it should coincide with experimental 
results for any sample. 

The expectation that the assumption of self-averaging is valid is based on 
the law of large numbers. The theorem says that the 
mean of independent random variables is equal to the expectation value 
with probability one. 
However, disordered systems such as amorphous semiconductors are 
strongly disordered, i.e., the expectation value of 
the random variables that describe structural disorder diverge. 
For example, the dispersive transport has been 
successfully explained by strongly disordered random variables 
\cite{Scher91}. In the theory of the dispersive transport, 
the transport is modeled by a random walk 
model on the disordered lattice where the dwell times of carriers 
at lattice sites are strongly disordered random variables. 

The distribution of strongly disordered random variables is broad. 
Hence, properties of limited areas of samples like the maximal dwell time 
govern the macroscopic properties like the macroscopic mobility. 
One simple example is the finite contribution of the maximal term of a set of 
strong disordered random variables to the sum of the set 
(see Eq.\ (\ref{max_strong_disorder}) in appendix.). 
The properties of the limited areas with principal contribution fluctuate 
from one sample to another. Hence, the macroscopic properties of such 
materials may fluctuate. Thus, the assumption of self-averaging should 
be tested for its validity in strongly disordered systems. 
In fact, there are several reports on absence of self-averaging in 
transport phenomena in disordered media: 
Ref.\ 3 for mean square displacement, 
ref.\ 4 for the mean first passage time in the Sinai model.

In this paper, we study theoretically self-averaging properties of 
physical quantities obtained in a boundary perturbation experiment. 
The experimental technique was introduced recently to measure the
optoelectrical properties of amorphous semiconductors \cite{Jongh96}. 
In this experimental method, an oscillatory perturbation with laser
light is applied at one end of a sample and a response of the photocurrent
from the other end is measured. 
We use as a generic model of the boundary perturbation experiment 
a random walk model in a one dimensional lattice with a boundary condition 
oscillating in time. The potential surface that carriers in the amorphous 
semiconductors feel is rugged due to structural disorder and it is called 
a rugged energy landscape. Consequently, 
the jump rates of the carriers are random variables and its probability 
distribution characterizes the disorder. 
It is known that the random jump rates obeys the power-law distribution 
with a negative exponent \cite{Pfister78} and it implies that the energy
landscapes of the disordered media are extremely rugged. 

It is shown that the complex admittance obtained in the experiment 
is not self-averaging when the jump rates $w_i$ are random variables
distributed by the power-law distribution with the negative exponent. 
More precisely, the frequency-dependence of the disorder-averaged 
admittance $\langle \chi \rangle$ disagrees with that of the admittance
$\chi$ of any sample. It implies that the Cole-Cole plot of $\langle
\chi \rangle$ shows a different shape from that of the Cole-Cole plots
of $\chi$ of each sample. The condition for absence of self-averaging is
investigated with a toy version of the complex admittance. 
The complex admittance for higher dimensional media is investigated and shown 
to be non-self-averaging. 
 
The rest of this paper is organized in the following way: 
In Sec.\ \ref{sec:model}, the random walk model used in the present study 
is explained. In terms of random walk, the complex admittance is 
related to a more physically relevant quantity to describe the transport 
in a disordered medium in Sec.\ \ref{sec:cafptd}. 
In Sec.\ \ref{sec:absence}, absence of self-averaging for the 
complex admittance of one-dimensional media is shown. 
In Sec.\ \ref{comp_admitt_section}, absence of self-averaging is displayed in 
the Cole-Cole plot impressibly. In order to understand the condition for 
absence of self-averaging, a toy model is studied with 
the extended central limit theorem in Sec.\ \ref{toy_section}. 
Absence of self-averaging for higher dimensional disordered media is 
shown in the similar way for the one-dimensional case in Sec.\ \ref{sec:h-d}.
Discussions and conclusions including consideration of possibilities that 
non-self-averaging properties are observed in real experiments 
are given in Sec.\ \ref{sec:conclusion}.
A part of this work has been reported in the brief note 
\cite{Kawasaki00} without 
consideration based on the probability distribution of the 
normalized mean first passage time, study of the toy model 
and the results for higher dimensional media.

\section{A generic model of the boundary perturbation experiment}
\label{sec:model}

Response of a system to an external field is a standard tool to
be utilized in the study of condensed matter.
Recently, the intensity modulated photocurrent spectroscopy 
\cite{Jongh96} has been introduced to investigate transport properties
in amorphous semiconductors. In the experiment, 
the electrode is illuminated with a laser and the incident light intensity is 
harmonically modulated at frequency $\omega$. 
The light absorbed by an optical transition 
generates the photocurrent. The light intensity absorbed in the electrode 
consists of a background intensity and an oscillating component
with small amplitude $\phi(\omega)$, which respectively gives rise to a 
steady state photocurrent and a harmonically varying photocurrent 
$\triangle i(\omega)$. 
Since the harmonically varying photocurrent $\triangle i(\omega)$ may
show a phase shift with respect to the absorbed light flux
$\phi(\omega)$, 
the optoelectrical admittance $\triangle i(\omega)/\phi(\omega)$ becomes 
a complex number.

Since in this experiment an oscillatory perturbation is applied at one end 
of a system and a response is measured, 
the experimental technique belongs to a 
generic method called the boundary perturbation method (BPM). 
In the presence of a periodically forced boundary condition on one end of 
the system, the output from the other end of the system is in proportion 
to the perturbation in the linear response regime and the 
proportionality constant is 
called the admittance. The frequency dependence of the admittance contains 
various information of the dynamics of the system.
The BPM is expected to provide useful information on transport properties of 
disordered media.

\begin{figure}
\includegraphics[width=10cm,keepaspectratio]{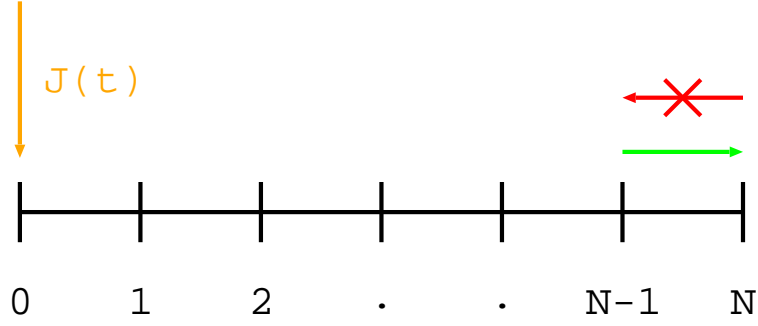}
\caption{The structure of the one-dimensional lattice segment. 
The oscillatory current perturbation $J(t)$ is introduced at the site $0$. 
The site $N$ is assumed to be a sink.}
\label{chain}
\end{figure}

We introduce as a generic model of the BPM a random walk model in 
one-dimensional disordered lattice segment of $N+1$ sites 
with an oscillatory input current (see Fig.\ \ref{chain}). 
The lattice sites are denoted by integers, $n=0,1,\ldots,N$.
The dynamics of a random walking particle can be described by 
a master equation for the probability $P_n(t)$ that the particle is 
at the site $n$ at time $t \geq 0$.
The master equation for the model is written as
\begin{equation}
\frac{d P_n(t)}{dt}=w_{n \ n-1}P_{n-1}(t)-(w_{n-1 \ n}+
w_{n+1 \ n})P_{n}(t)+w_{n \ n+1}P_{n+1}(t),
\label{master1}
\end{equation}
where $w_{m \ n}$ denotes the random 
jump rate of a particle from the site $n$ to the site $m$.
Here, we assume that the random walking particle jumps only to the 
nearest neighbor sites. 
The probability distribution of jump rates characterizes the disordered 
medium. 
We introduce a perturbation at the left end, 
so that the equation for $P_0(t)$ 
is given by 
\begin{equation}
\frac{d P_0(t)}{dt}=-w_{1 \ 0}P_{0}(t)+w_{0 \ 1}P_1(t)+J(t).
\label{master2}
\end{equation}
$J(t)$ is the oscillatory current perturbation at the site $0$. 
The right end of the system is assumed to be a sink and the equation for 
$ P_{N-1}$ is given by
\begin{equation}
\frac{d P_{N-1}(t)}{dt}=w_{N-1 \ N-2}P_{N-2}(t)-(w_{N-2 \ N-1}+w_{N \ N-1})
P_{N-1}(t).
\label{master3}
\end{equation} 
Since the current perturbation $J(t)$ oscillates in time around 
a positive average with 
the amplitude $\triangle J$, the response of the output current 
$w_{N \ N-1} P_{N-1}(t)$ oscillates around its stationary state 
with the amplitude $w_{N \ N-1} \triangle P_{N-1}$ at the same frequency 
with a phase-shift. 
The admittance is defined by the ratio of these two amplitudes
\begin{equation}
\chi_N(\omega)\equiv \frac{w_{N \ N-1} \triangle P_{N-1}}{\triangle J}.
\label{adm}
\end{equation}
The system of equations to determine 
the amplitudes $\triangle P_n$ of oscillation of the probabilities 
is derived from the master equation Eqs. (\ref{master1}),
(\ref{master2}) and (\ref{master3}); 
\bea
i \omega \triangle P_n & = & w_{n \ n-1} \triangle P_{n-1}-
(w_{n-1 \ n}+w_{n+1 \ n})\triangle P_n
\nonumber \\
& & +w_{n \ n+1}\triangle P 
\ \ \mbox{for $n=2,3,\ldots,N-2$},\nonumber \\
i \omega \triangle P_0 & = & 
-w_{1 \ 0}\triangle P_0+w_{0 \ 1}
\triangle P_1+\triangle J, \nonumber \\
i\omega \triangle P_{N-1} & = & w_{N-1 \ N-2}\triangle 
P_{N-2}-
(w_{N-2 \ N-1}+w_{N \ N-1})\triangle P_{N-1}.
\label{amplitude_eq}
\eea

\section{The complex admittance and the first passage time distribution}
\label{sec:cafptd}

We show below that the admittance can be related to the first passage time 
distribution $F_{N \ 0}(t)$ (FPTD), which is the probability density that 
the particle which starts at the site $0$ at time $0$ 
arrives for the first time at the site $N$ at time $t$.
Since the site $N$ is a sink, the FPTD is given by the output current 
from the site $N$ when there is no input current 
and the particle starts from the site $0$ at time $0$, 
i.e. 
\be
F_{N \ 0}(t)=w_{N \ N-1}P_{N-1}(t). 
\label{fptd}
\ee
The Fourier-Laplace transform of the 
master equation for the case is written as 
\bea
i \omega \tilde{P}_n(i\omega) & = & w_{n \ n-1} \tilde{P}_{n-1}(i\omega)-
(w_{n-1 \ n}+w_{n+1 \ n})\tilde{P}_n(i\omega)
\nonumber \\
& & +w_{n \ n+1}\tilde{P}(i
\omega) \ \ \mbox{for $n=2,3,\ldots,N-2$},\nonumber \\
i \omega \tilde{P}_0(i\omega) & = & -w_{1 \ 0}\tilde{P}(i\omega)_0+w_{0 \ 1}
\tilde{P}_1(i\omega)+1, \nonumber \\
i\omega \tilde{P}_{N-1}(i\omega) & = & w_{N-1 \ N-2}\tilde{P}_{N-2}(i\omega)-
(w_{N-2 \ N-1}+
\nonumber \\
& & w_{N \ N-1})\tilde{P}_{N-1}(i\omega).
\eea
This set of equations is identical to Eq.\ (\ref{amplitude_eq}) 
divided by $\triangle J$. 
It means that $\triangle P_n/\triangle J = \tilde{P}_n(i\omega)$ and 
hence the Fourier-Laplace transform of 
the FPTD $\tilde{F}_{N \ 0}(s=i \omega)$ 
is identical to $ w_{N \ N-1} \triangle P_{N-1}/ \triangle J$.
Thus, from Eqs.\ (\ref{adm}) and (\ref{fptd}), we conclude that 
the admittance is equal to the Fourier-Laplace 
transform of the FPTD;
\be
\chi_N(\omega)=\tilde{F}_{N \ 0}(i\omega).
\ee

We make the low-frequency expansion of the admittance to see 
the behavior near the static limit.
Since the admittance is given by the Fourier-Laplace transform of the FPTD, 
the admittance at zero frequency equals to unity due to normalization 
of the FPTD; 
\be
\tilde{F}_{N \ 0}(0)=\int_0^{\infty}dt F_{N \ 0}(t)=1.
\ee
The mean first-passage time (MFPT) $\bar{t}_N$, which is the first moment 
of the FPTD, is given by $\bar{t}_N=i \ d\chi(\omega = 0)/d\omega$.
Thus, the low-frequency expansion of the admittance is given by 
\begin{equation}
\chi_N(\omega)=1-i \ \bar{t}_N \ \omega+O(\omega^2). \label{adm2}
\end{equation}
The MFPT is given by \cite{Murthy89}
\begin{equation}
\bar{t}_N=\sum^{N-1}_{k=0} \frac{1}{w_{k-1 \ k}}+
\sum^{N-2}_{k=0}\frac{1}{w_{k+1 \ k}}\sum^{N-1}_{i=k+1} \prod^{i}_{j=k+1}
\frac{w_{j-1 \ j}}{w_{j+1 \ j}}. \label{mfpt}
\end{equation}

\section{Absence of self-averaging for the complex admittance}
\label{sec:absence}

Since we are interested in the complex admittance for amorphous
semiconductors, we employ the power law probability distribution of the
random jump rates Eq.\ (\ref{jrdist}), which is valid for amorphous
semiconductors \cite{footnote1}:
\begin{equation}
 \rho(w) = \left\{ \begin{array}{ll}
			\alpha w^{\alpha-1} & \mbox{if $0<w<1$} \\
			0                  & \mbox{otherwise.}
		  \end{array}
	  \right.  \label{jrdist}
\end{equation}

There is possibility of the absence of self-averaging when disorder of the 
random dwell time at a site is strong as suggested in Sec.\ 
\ref{sec:introduction}. 
Since the dwell time is proportional to inverse of the jump rate, 
we are interested in the case where the first inverse moment of the 
random jump rate $\langle 1/w \rangle$ diverges, i.e. 
$0<\alpha\leq 1$.
It is important to note that the inequality $0<\alpha<1$ holds 
for amorphous semiconductors \cite{Pfister78}. 

\subsection{The probability distribution of the MFPT}

In this subsection, we analyze the 
probability distribution of the MFPT Eq.\ (\ref{mfpt}). 
Since the MFPT gives the coefficient of the low frequency expansion of 
the admittance, the results of the analysis on the MFPT make it possible 
to determine whether the admittance is distributed with finite variance, 
i.e. non-self-averaging, or not. We show below that the 
complex admittance Eq.\ (\ref{adm2}) is non-self-averaging when $0<\alpha<1$.

In order to analyze Eq.\ (\ref{mfpt}), we employ the
random trap model \cite{general interest}, 
where the jump rates $w_{ij}$ depends only on $j$. 
For the random barrier model, 
where the jump rates have the symmetry such 
that $w_{ij}=w_{ji}$, the {\it quantitatively} same results 
for self-averaging properties are obtained. 
We discuss below the probability distribution function of the
normalized MFPT for the three cases of the value of $\alpha$. 
It is important to note that the three cases exhaust all possibilities
of the value of $\alpha$. The choice of the normalization constants is
suggested by the extended central limit theorem \cite{Feller66}.
The theorem is summarized in appendix \ref{appendix:eclt}.

\begin{enumerate}
\item The case when $0<\alpha<1$ is considered. It is numerically shown that 
      the probability distribution function of 
      the normalized MFPT defined as $\bar{t}_N/N^{1/\alpha+1}$ 
      converges
      to a distribution function with non-zero variance in the limit 
      $N\rightarrow\infty$. Namely the normalized MFPT, 
      $\lim_{N\rightarrow\infty} \bar{t}_N/N^{1/\alpha+1}$, is a random 
      variable even for the infinitely long chain. 
      In addition, the distribution function behaves
      asymptotically as $\rho(x) \sim x^{-\gamma} (1<\gamma <2)$ for large 
      $x$. It is shown in Sec.\ \ref{sec:pdf-mfpt-dd} that $\gamma = \alpha+1$.
	\begin{figure}
	\includegraphics[width=10cm,keepaspectratio]{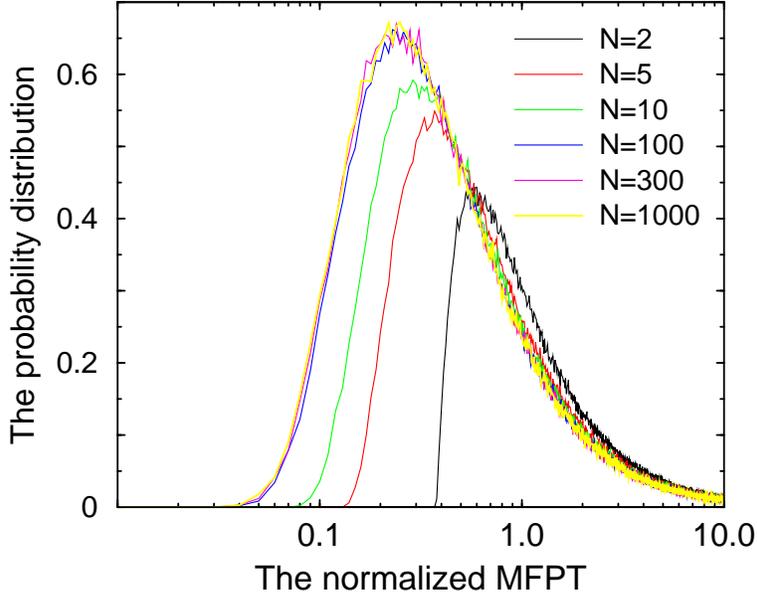}
	\caption{The probability distribution function of the normalized MFPT 
	 $\bar{t}_N/N^{1/\alpha+1}$ for
	 the random trap model where the probability distribution function of
	 the random jump rates is power-law Eq.\ (\ref{jrdist}) 
	with exponent $\alpha=1/2$. 
	The distribution is computed from 500000 samples of disordered chains 
	of the length $N=2, 5, 10, 100, 300, 1000$. 
	It is clearly seen that the probability 
	distribution functions lie on a same curve when the length of lattice
	 $N$ is sufficiently large.}
	\label{dist_mfpt}
	\end{figure}
      Figure \ref{dist_mfpt} shows the probability distribution function of 
      $\bar{t}_N/N^{1/\alpha+1}$ when $\alpha=1/2$ in Eq.\
      (\ref{jrdist}). 
      It is clearly seen that the probability distribution functions 
      for sufficiently long chains lie on a same curve.
\begin{figure}
\includegraphics[width=10cm,keepaspectratio]{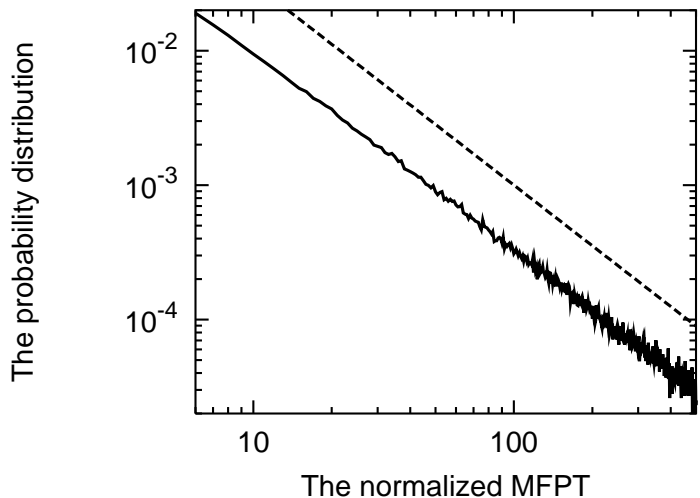}
\caption{The asymptotic behavior of the probability distribution
 function of the normalized MFPT computed from 5000000 samples. 
 The solid line denotes the
 distribution function when $N=1000$ and $\alpha=1/2$ 
 and the dashed line represents the
 function proportional to $x^{-1.5}$. One sees clearly that the
 distribution function behaves asymptotically as $\rho(x) \sim x^{-\gamma} (\gamma = 1.5 < 2).$}
\label{dist_mfpt3}
\end{figure}
      Figure \ref{dist_mfpt3} shows the tail of the probability distribution 
      which appears in Fig.\ \ref{dist_mfpt}. It behaves asymptotically as 
      $\rho(x) \sim x^{-\gamma} \ (\gamma = 1.5<2)$.
\item When $\alpha=1$, it is numerically shown that the probability
      distribution function of the normalized MFPT defined as
      $\bar{t}_N/N^2-b_N/2$ converges to a distribution function with
      non-zero variance in the limit $N\rightarrow\infty$. Here, the
      centering constant $b_N$ is defined as 
      \begin{equation}
       b_N \equiv N \int^{\infty}_1 dx \sin(x/N) x^{-2}.
	\label{b_N}
      \end{equation}
      In appendix \ref{sec_centering_constant}, we show that $b_N$ diverges as
      $N\rightarrow\infty$.
\begin{figure}
\includegraphics[width=10cm,keepaspectratio]{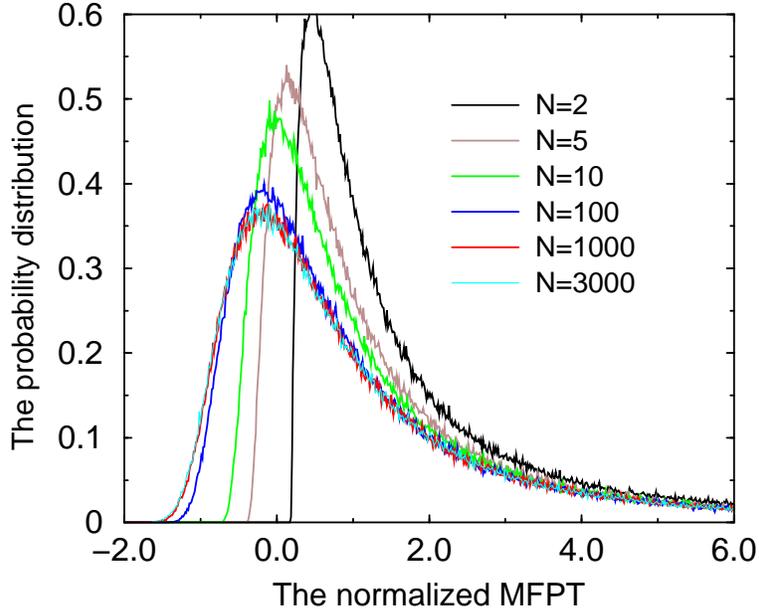}
\caption{The probability distribution function of the normalized MFPT 
 $\bar{t}_N/N^{2}-b_N/2$ for
 the random trap model where the probability distribution function of
 the random jump rates is power-law Eq.\ (\ref{jrdist}) 
with exponent $\alpha=1$. 
The distribution is computed from 500000 samples of disordered chains 
of the length $N=2, 5, 10, 100, 300, 1000, 3000$. 
It is clearly seen that the probability 
distribution functions lie on a same curve when the length of lattice
 $N$ is sufficiently large.}
\label{dist_mfpt.1.0}
\end{figure}
      Figure.\ \ref{dist_mfpt.1.0} shows the
      probability distribution function of the normalized MFPT when
      $\alpha=1$. It is clearly seen that the probability distribution
      functions lie on a same curve when the length of lattice $N$ is
      sufficiently large. 
\item When $1<\alpha$, it is also shown numerically that the probability 
      distribution function of the normalized MFPT defined as
      $\bar{t}_N/N^2$ converges to the Dirac delta function with the
      support at $\langle 1/w \rangle /2$. 
      Figure \ref{dist_mfpt2} shows the probability
      distribution function of $\bar{t}_N/N^2$ when $\alpha=1.5$. 
      One sees that the probability 
      distribution concentrates on the point 
      $\langle 1/w \rangle/2=1.5$ when the length $N$ is large. 
\begin{figure}
\includegraphics[width=10cm,keepaspectratio]{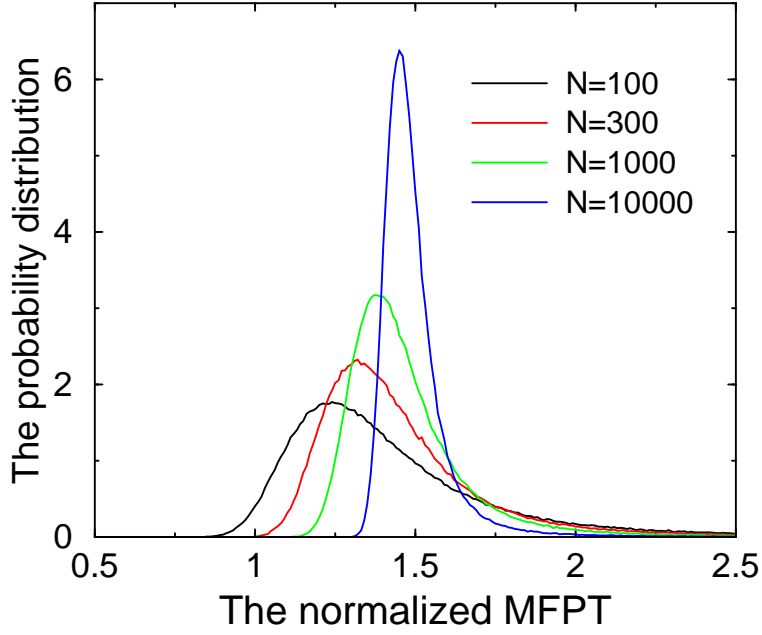}
\caption{The probability distribution function of the normalized MFPT 
 $\bar{t}_N/N^{2}$ when $\alpha=1.5$. 
It is computed from 500000 samples of disordered chains of length 
$N=100, 300, 1000, 10000$. It is seen clearly that the probability 
distribution concentrates on the point $\langle 1/w \rangle/2=1.5$  
when the length of lattice $N$ is sufficiently large.}
\label{dist_mfpt2}
\end{figure}
\end{enumerate}

\subsection{Absence of self-averaging in the low frequency behavior of the 
complex admittance}

By using the results for the probability distribution of the normalized
MFPT, we discuss below frequency dependence of the complex admittance
Eq.\ (\ref{adm2}) for the three cases of the value of $\alpha$:
$0<\alpha<1$, $\alpha=1$ and $1< \alpha$. 

\begin{enumerate}
\item When $0<\alpha<1$, it is shown that the complex admittance is
      non-self-averaging. In order to see frequency dependence of the
      admittance, the long chain limit ($N\rightarrow\infty$) should be
      taken with keeping $\bar{t}_N\omega$ finite. Otherwise, the
      admittance becomes a trivial constant $0$ or $1$. Since it 
      was shown that the normalized MFPT $\bar{t}_N/N^{1/\alpha+1}$ is
      finite with probability one, the long chain limit should be taken
      with fixing $\omega N^{1/\alpha+1}$ finite. Thus, from Eq.\
      (\ref{adm2}), the admittance for the infinitely long chain is given
      by 
      \be
      \lim_{{N\rightarrow\infty} \atop{\; \tilde{\omega}=\omega 
      N^{1/\alpha+1}}}
      \chi_N(\omega) = 1-i \ \tilde{t} \ \tilde{\omega}+\ldots
      \label{long_adm}
      \ee
      where $ \tilde{t}$ is the normalized MFPT defined as $ \tilde{t}\equiv 
      \lim_{N\rightarrow\infty}\bar{t}_N/N^{1/\alpha+1}$ and 
      $\lim_{{N\rightarrow\infty} \atop{\; \tilde{\omega}=\omega 
      N^{1/\alpha+1}}}$
      means the long chain limit with fixing 
      $\omega N^{1/\alpha+1}$ at the finite value $\tilde{\omega}$. 
      Equation (\ref{long_adm}) indicates that 
      the admittance for the infinitely long chain is a function of 
      the normalized MFPT $ \tilde{t}$ and the scaled frequency $
      \tilde{\omega}$. From our result for the probability distribution
      of the normalized MFPT 
      when $0<\alpha<1$, $\tilde{t}$ is a random
      variable. Thus, the admittance is also a random variable and hence the
      admittance is non-self-averaging. 
\item When $\alpha=1$, it is shown below that the admittance is self-averaging 
      in the low frequency range. 
      By introducing the normalized MFPT $\tilde{t}_N$ defined as 
      \be
      \tilde{t}_N \equiv \frac{\bar{t}_N}{N^2}-\frac{b_N}{2}, 
      \ee
      $\bar{t}_N \omega$ is rewritten as 
      \be
      \bar{t}_N\omega = N^2 b_N \omega (\frac{2}{b_N}\tilde{t}_N+1)/2.
      \ee
      It was shown in Fig.\ \ref{dist_mfpt.1.0} 
      that the normalized MFPT $\tilde{t}_N$ 
      is finite with probability one. 
      Thus, in order to see frequency dependence, 
      the long chain limit should be taken with keeping 
      the scaled frequency $\tilde{\omega}\equiv N^2 b_N \omega/2$
      finite. Since $b_N$ diverges as $N\rightarrow\infty$, the
      complex admittance for the infinitely long chain is given by 
      \begin{eqnarray}
       \lim_{{N\rightarrow\infty} \atop{\; \tilde{\omega}=N^2 b_N \omega/2}} 
       \chi_N(\omega) & = & 
       \lim_{{N\rightarrow\infty} \atop{\; \tilde{\omega}=N^2 b_N \omega/2}} 
       1-i N^2 b_N \omega/2 (\frac{2}{b_N}\tilde{t}_N+1)+\ldots 
       \nonumber \\
       & = & 1-i \tilde{\omega}+\ldots
       \label{adm.alpha=1}
      \end{eqnarray}
      Since Eq.\ (\ref{adm.alpha=1}) has no sample-dependence, 
      the complex admittance is self-averaging in the low frequency
      region. 
\item When $1<\alpha$, it is shown below that the complex admittance is
      self-averaging. In this case, it was shown that the scaled
      MFPT $\bar{t}_N/N^2$ is equal to $\langle 1/w \rangle /2$ in the
      limit $N\rightarrow\infty$. Thus, the admittance for the
      infinitely long chain is given by 
      \be
      \lim_{{N\rightarrow\infty}\atop{\; \tilde{\omega}=\omega N^2}} 
      \chi_N(\omega)=1-\frac{1}{2}\left\langle \frac{1}{w} \right\rangle 
      \tilde{\omega}+\ldots
      \label{adm.alpha.g.1}
      \ee
      It means that the complex admittance is self-averaging.
\end{enumerate}

From Eq.\ (\ref{long_adm}), the low-frequency expansion of 
the disorder-averaged admittance is given by 
\be
\langle 
\lim_{{N\rightarrow\infty} \atop{\; \tilde{\omega}=\omega N^{\beta}}} 
\chi_N(\omega) \rangle = 
1-i \ \langle \tilde{t} \rangle \ \tilde{\omega}+\ldots
\label{ave_adm_low}
\ee
When $0<\alpha<1$, the distribution function
of the normalized MFPT $\tilde{t}$ behaves asymptotically as
$\rho(x)\sim x^{-\gamma}$. 
It is shown later in Sec.\ \ref{sec:pdf-mfpt-dd} that 
$\gamma = \alpha+1$ and it means that $1<\gamma<2$. 
Hence, the expectation value $\langle \tilde{t}
\rangle$ of the normalized MFPT diverges. 
Since it implies that the coefficient of the first order of low 
$\tilde{\omega}$ expansion diverges, 
the disorder-averaged admittance is a non-analytic function of 
$i \tilde{\omega}$ which must behave as 
\be
\langle 
\lim_{{N\rightarrow\infty} \atop{\; \tilde{\omega}=\omega N^{\beta}}} 
\chi_N(\omega) \rangle -1 \sim (i\tilde{\omega})^{\mu}
\label{achi}
\ee
where $0<\mu<1$.

\begin{figure}
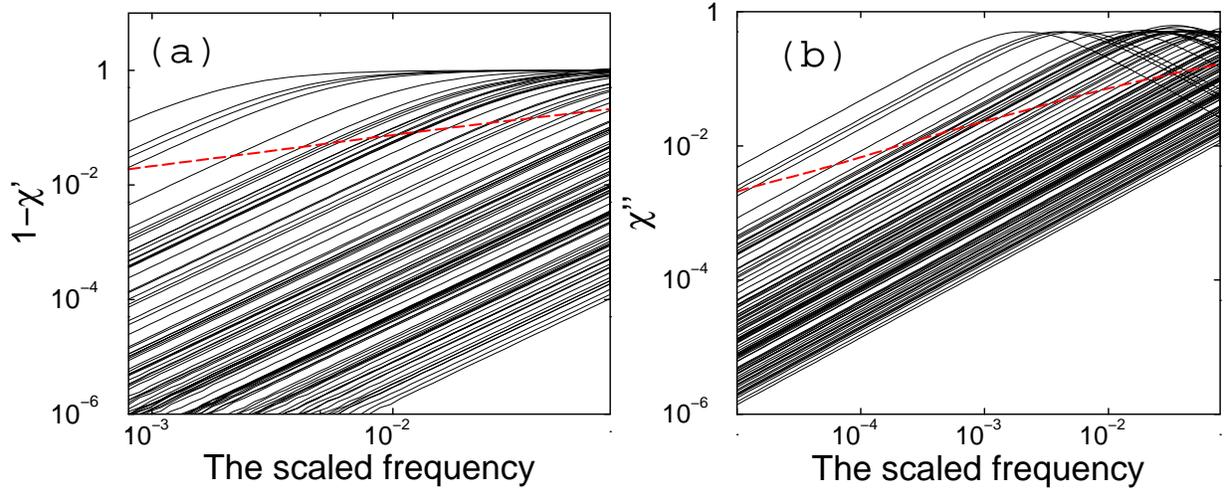

\includegraphics[width=8cm,keepaspectratio]{Fig6a.eps} 
\includegraphics[width=8cm,keepaspectratio]{Fig6b.eps}
\caption{The frequency dependence of the real part (a) and 
the imaginary part (b) of the complex admittances 
when the exponent of the probability distribution of jump rates 
is $\alpha=1/2$ and the length of a chain is $N=20$.
Solid lines represent the admittances for each of 100 samples.
The dashed line represents the admittance averaged over 5000 samples.
Though the admittance of each sample is proportional to 
the scaled frequency $\tilde{\omega}$,
the averaged admittance is proportional to $\sqrt{\tilde{\omega}}$ 
($\mu \simeq 1/2$).}
\label{low-frequency}
\end{figure}

\begin{figure}
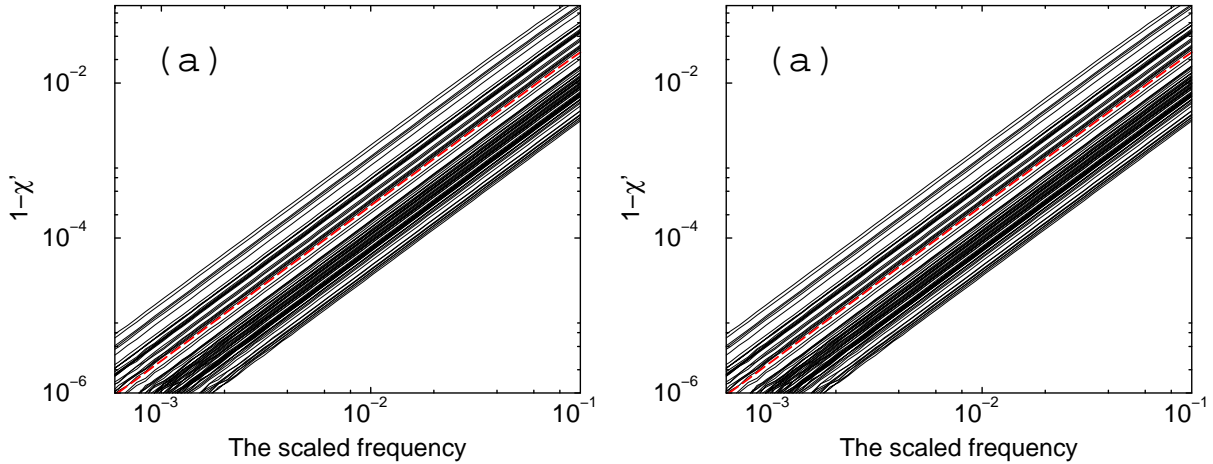

\includegraphics[width=8cm,keepaspectratio]{Fig7a.eps} 
\includegraphics[width=8cm,keepaspectratio]{Fig7b.eps}
\caption{The frequency dependence of the real part (a) and 
the imaginary part (b) of the complex admittances 
when $\alpha=1.0$ and the length of a chain is $20$.
Solid lines represent the admittances for each of 100 samples.
The dashed line represents the admittance averaged over 5000 samples.
Both the admittance of each sample and the averaged admittance are 
proportional to the scaled frequency $\tilde{\omega}$.}
\label{low-frequency3}
\end{figure}

\begin{figure}
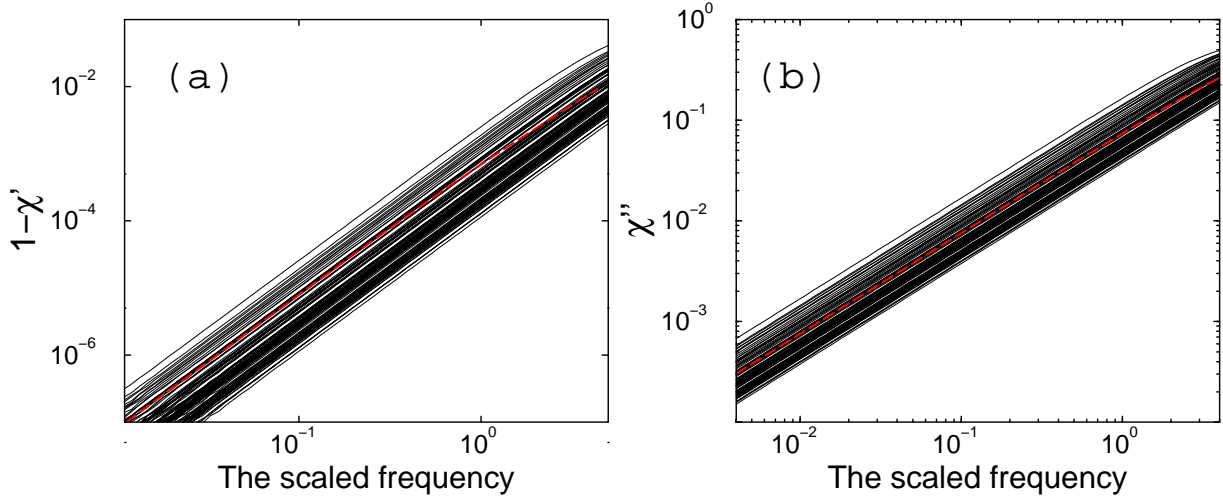

\includegraphics[width=8cm,keepaspectratio]{Fig8a.eps} 
\includegraphics[width=8cm,keepaspectratio]{Fig8b.eps}
\caption{The frequency dependence of the real part (a) and the 
imaginary part (b) of the complex admittances 
when $\alpha=1.5$ and the length of a chain is $20$.
Solid lines represent the admittances for each of 100 samples.
The dashed line represents the admittance averaged over 5000 samples.
Both the admittance of each sample and the averaged admittance are 
proportional to the scaled frequency $\tilde{\omega}$.}
\label{low-frequency2}
\end{figure}

In order to test the foregoing observation, we numerically solved 
Eq.\ (\ref{amplitude_eq}) for one-dimensional chain of $20$ sites. 
The jump rate $w_{i \ j}$ depends only on $j$ (the random trap model) 
and the jump rates are generated from the power-law distribution 
Eq.\ (\ref{jrdist}). 
Figure \ref{low-frequency} shows the low-frequency behavior of 
the real part and the imaginary part of the admittances when
$\alpha=1/2$. 
One clearly sees that the low-frequency behavior of the 
disorder-averaged admittance is completely different from 
the admittance for any of the samples. 
On the other hand, figures \ref{low-frequency3} and \ref{low-frequency2} 
show the low-frequency behavior 
for $\alpha=1.0$ and for $\alpha=1.5$. 
One clearly sees that 
the admittances for individual samples and their 
disorder-averaged admittance are both proportional to $\tilde{\omega}$.
Although sample-dependence is observed, the sample-dependence is considered 
as the finite-size effect or effects from the terms of $O(\omega^2)$.

\section{Non-self-averaging in the Cole-Cole plot}
\label{comp_admitt_section}

In the literature of experiments of the BPM \cite{Jongh96}, 
the Cole-Cole plot of the admittance is employed for the analysis of 
experimental results. 
It is a parametric plot of the imaginary part of the admittance 
against the real part. 
\begin{figure}
\includegraphics[width=10cm,keepaspectratio]{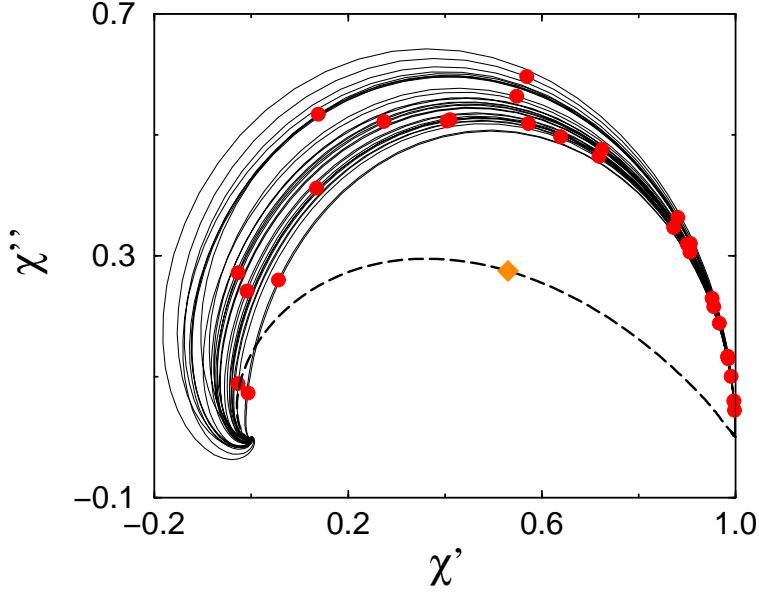}
\caption{The Cole-Cole plots of the admittances 
when the exponent of the probability distribution of jump rates is 
$\alpha=1/2$ and
$N=100$. Solid lines represent the admittances for each of 30 samples.
The dashed line represents the admittance averaged over 5000 samples.
The red solid circles represent the admittance of each sample 
at $\tilde{\omega} = 0.5$.
Although the red solid circles scatter outside the Debye semi-circle, 
the yellow diamond which denotes 
the averaged admittance at $\tilde{\omega}=0.5$ 
is inside of the Debye semi-circle.}
\label{Cole-Cole}
\end{figure}
Figure \ref{Cole-Cole} shows the Cole-Cole plot of $ \chi $ and 
$\langle \chi \rangle$ obtained by solving Eq.\ (\ref{amplitude_eq}) 
when $\alpha=1/2$ in Eq.\ (\ref{jrdist}).
One sees that the shape of the Cole-Cole plot of the disorder-averaged 
admittance is completely different from the shape of the Cole-Cole plot 
for each of samples. 
The boundary of the region where the Cole-Cole plot of the admittance of 
each sample scatters is the Debye semi-circle, 
whose center is at $(1/2,0)$ and the radius is $1/2$. 
In Fig.\ \ref{Cole-Cole}, solid circles denote the values of the 
admittance at a given frequency and a diamond is the average over 
these values. 
Since the admittance of each of samples at the same frequency (a solid 
circle) scatters around the Debye semi-circle, the admittance averaged at the 
same frequency (the diamond) comes inside the Debye semi-circle.

In order to analyze this non-self-averaging property of the admittance in 
the Cole-Cole plot, we prove rigorously that the Cole-Cole plots of each 
samples must lie outside the Debye semi-circle. 
We first note that the inverse of the admittance satisfies the 
following recursive relation:
\begin{equation}
\chi_{n}^{-1}(\omega)=\left(\frac{i \omega}{w_{n+1 \ n}}+1+
\frac{w_{n-1 \ n}}{w_{n+1 \ n}}\right) \chi^{-1}_{n-1}(\omega)-
\frac{w_{n-1 \ n}}{w_{n+1 \ n}} \chi^{-1}_{n-2}(\omega),
\label{rec}
\end{equation}
where $\chi_{n}^{-1}(\omega)$ is the inverse of the admittance 
for a chain of length $n$. 
The recursive Eq.\ (\ref{rec}) is derived from the
following two equations for the FPTD obtained from its definition:
$\tilde{F}_{i+1 \ i}(s)=\tilde{\psi}_{i+1 \ i}(s)+\tilde{\psi}_{i-1 \ i}(s)
\tilde{F}_{i+1 \ i-1}(s)$ and 
$\tilde{F}_{i+1 \ i-1}(s)=\tilde{F}_{i+1 \ i}(s) \tilde{F}_{i \ i-1}(s)$,
where $\tilde{F}_{i \ j}(s)$ denotes the Laplace transform of the FPTD
from site $j$ to site $i$ and $\tilde{\psi}_{i \ j}(s)$ denotes the Laplace 
transform of the waiting time distribution for the jump 
from site $j$ to site $i$.
The recursive relation proves inductively 
that the inverse of the admittance obeys 
the following four inequalities in the frequency region $[0, \omega_n]$, 
where $\omega_n$ is the positive smallest zero of the real part of 
$\chi_n^{-1}$ and is 
equal to the positive smallest zero of ${\chi_n}'$:
\begin{equation}
\left\{ \begin{array}{l}     {\chi_n^{-1}}'' > 0 \\
	                     {\chi_{n-1}^{-1}}' > 0 \\
	                     {\chi_n^{-1}}'' > {\chi_{n-1}^{-1}}'' \\
	                     {\chi_n^{-1}}' < {\chi_{n-1}^{-1}}'.
	\end{array} 
\right.
\end{equation}
Since the second inequality shows $\omega_n \leq \omega_{n-1}$,
the fourth inequality implies 
${\chi_n^{-1}}' < {\chi_{n-1}^{-1}}' < \cdots < {\chi_1^{-1}}'=1$ 
in the region $[0,\omega_n]$. 
Since ${\chi_n^{-1}}'={\chi_n}'/({{\chi_n}'}^2+{{\chi_n}''}^2)$, the
inequality ${\chi_n^{-1}}' <1$ means $({\chi_n}'-1/2)^2+{{\chi_n}''}^2 > 
(1/2)^2$. 
Thus, the Cole-Cole plot of the 
admittance of each sample can exist only 
{\em outside the Debye semi-circle} when the 
frequency is smaller than the positive smallest zero of ${\chi_n}'$.

On the other hand, we can show in the following way that the
Cole-Cole plot of the averaged admittance is located  
{\em inside the Debye semi-circle} when $0<\alpha<1$.
We consider the angle $\theta$ between the tangent of the Cole-Cole plot at 
(1, 0) and the horizontal axis. From Eq.\ (\ref{achi}), $\theta=\mu \pi/2$, 
which is less than $\pi/2$ since $\mu < 1$, is obtained. 
However the angle $\theta$ for the Debye semi-circle is equal to $\pi/2$.
It means that the Cole-Cole plot of 
the averaged admittance exists inside the Debye semi-circle 
when $0 < \alpha < 1$.
Since the Cole-Cole plot of the admittance of any sample 
exists outside the Debye semi-circle, 
the Cole-Cole plot clearly shows non-self-averaging property of the admittance.

\begin{figure}
\includegraphics[width=10cm,keepaspectratio]{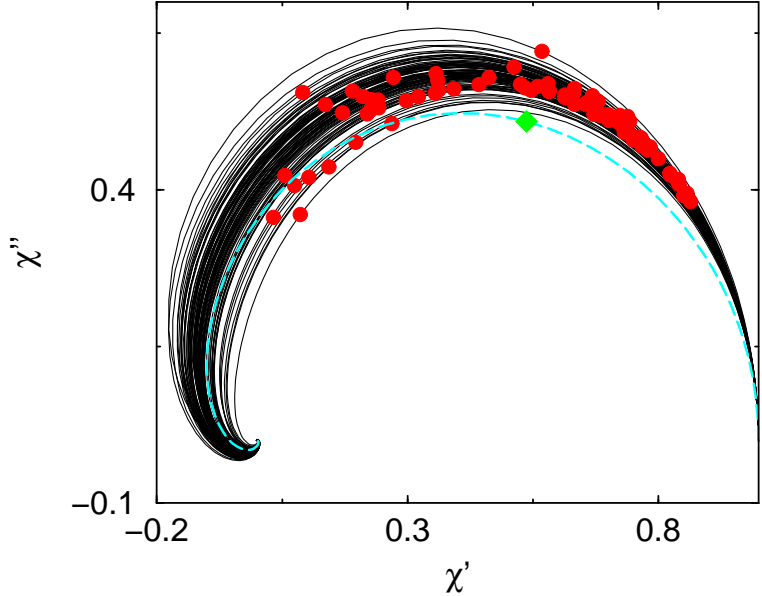}
\caption{The Cole-Cole plots of the admittances 
when the exponent of the probability distribution of jump rates is 
$\alpha=1.0$ and
$N=100$. Solid lines represent the admittances for each of 30 samples.
The dashed line represents the admittance averaged over 5000 samples.
The red solid circles represent the admittance of each sample 
at $\tilde{\omega} = 0.68$. 
The green diamond which denotes 
the averaged admittance at $\tilde{\omega}=0.68$ exists at different position 
from that of the red solid circles.}
\label{Cole-Cole.1.0}
\end{figure}

\begin{figure}
\includegraphics[width=10cm,keepaspectratio]{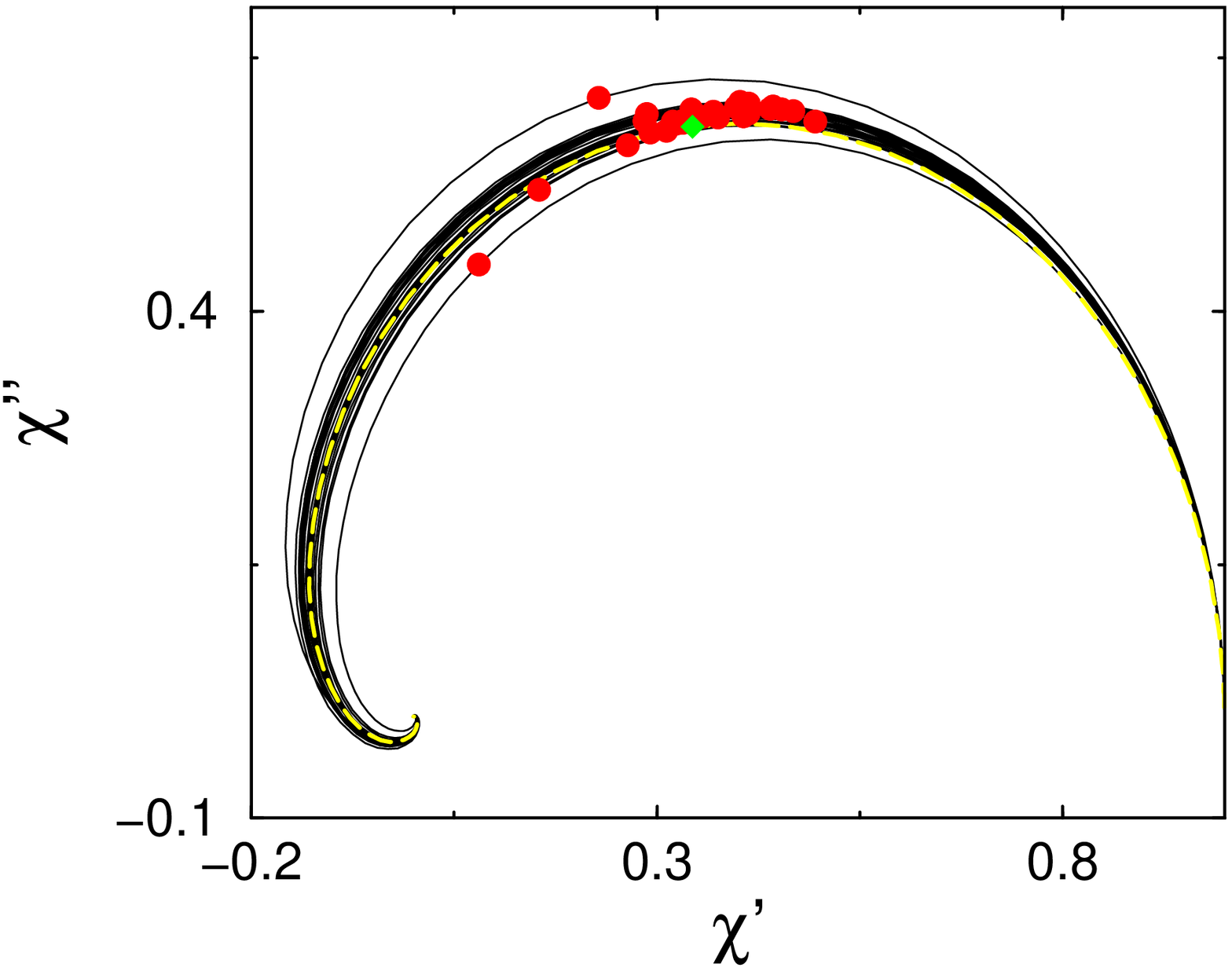}
\caption{The Cole-Cole plots of the admittances 
when $\alpha=1.5$ and 
$N=100$. Solid lines represent the admittances for each of 30 samples.
The yellow dashed line represents the admittance averaged over 5000 samples.
The red solid circles represent the admittance of each sample 
at $\tilde{\omega} = 1.0$.
The green diamond denotes the averaged admittance at 
$\tilde{\omega}=1.0$.
In contrast to the strongly disordered case Fig.\ \ref{Cole-Cole}, 
the circles concentrate in small region and the solid lines lie 
almost on the dashed line. }
\label{Cole-Cole2}
\end{figure}

When $\alpha \geq 1$, 
the angle $\theta$ of the Cole-Cole plot of the averaged admittance 
is $\pi/2$ and is equal to that of the Cole-Cole plot of each of samples 
as seen in Figs.\ \ref{Cole-Cole.1.0} and \ref{Cole-Cole2}. 
This property of the angle $\theta$ is derived from 
the regular behavior as $\langle \chi \rangle -1 \sim i 
\tilde{\omega}$ 
(see Eq.\ (\ref{adm.alpha=1}) and Eq.\ (\ref{adm.alpha.g.1}).). 

It is important to note that Fig.\ \ref{Cole-Cole.1.0} suggests that the 
complex admittance is non-self-averaging when $\alpha=1$. One sees that
the disorder-averaged complex admittance $\langle \chi \rangle$ exists
at different position from that of the complex admittance for each
samples. Although it was shown that the low frequency behavior is
self-averaging, it seems that the complex admittance at the finite
frequency is non-self-averaging. Consequently, we conclude that the
complex admittance is non-self-averaging when $0<\alpha\leq 1$.

\section{Understanding of non-self-averaging of the complex admittance 
with a toy model}
\label{toy_section} 
\newtheorem{theorem}{Theorem}[section]

In the previous section, 
we discussed absence of self-averaging for the admittance 
when disorder of the jump rates is strong. 
In this section, in order to consider the conditions for absence of 
self-averaging we introduce a simple model, 
which is a toy version of the complex admittance and 
can be analyzed rigorously.
This simple model will provide understanding of 
the reason why the admittance is non-self-averaging, 
since absence of self-averaging for the simple model is shown below with 
the same discussion as employed in the analysis of the complex
admittance. 

Analysis of the admittance is difficult because 
it does not consist of a mean of independent and 
identically distributed random variables. Hence, we consider 
a toy model of 
the admittance which is a simple function of 
a mean of independent and 
identically distributed random variables given by 
\be
\epsilon_N(\omega) \equiv \frac{1}{1+i \omega \sum^N_{i=1}(1/w_i)/N}.
\label{toy}
\ee
We call it the dielectric constant. 
It is obvious that the Cole-Cole plot of each sample is the Debye
semi-circle since the form of Eq. (\ref{toy}) is the same as that of the
dielectric constant of the Debye relaxation. 
In addition, 
since $|\epsilon_N(\omega)|$ is less than unity, $\epsilon_N(\omega)$ is 
finite for any set of the random jump rates ${w_i}$. 

In the analyses presented below, 
we use the following theorems of probability theory \cite{Feller66} 
summarized in appendix \ref{appendix:eclt};
Let $\{x_i\}$ be 
a set of independent and identically distributed random variables. 

\begin{theorem}[Law of large numbers] 
	   When the expectation value $ \langle x_i \rangle$ exists,
	   law of large numbers holds, i.e. 
	   \be
	   Prob\left( \left| \frac{\sum_{i=1}^N x_i}{N}-\langle x_j\rangle  
	   \right| >
	   \epsilon \right) \rightarrow 0 \ \mbox{as $N\rightarrow\infty$}.
	   \ee
\end{theorem}
\begin{theorem}[Corollary of the extended central limit theorem]
	   When the expectation value $\langle x_i\rangle$ diverges,
	   in order that the probability distribution of 
	   $\sum_{i=1}^N x_i/a_N-b_N$ converges as $N\rightarrow\infty$
	   it is necessary that the probability density $\rho(x_i)$ is
	   of the form 
	   \be
	   \rho(x) \sim x^{-\alpha-1} L(x)
	   \label{necessary_condition_convergence_probability_distribution}
           \ee
	   for $0<\alpha\leq 1$. Here, $L(x)$ is a slowly varying 
           positive function 
           which is precisely defined by Eq.\ (\ref{def_slow_variation}).
	   The normalization constant $a_N$ is chosen so that 
	   \be
	   N \frac{L(a_N)}{{a_N}^{\alpha}} \rightarrow C
	   \label{normalization_constant_1234}
	   \ee
	   as $N\rightarrow\infty$.
	   \begin{itemize}
	    \item When $0<\alpha<1$ in 
		  Eq.\ (\ref{necessary_condition_convergence_probability_distribution}), the probability distribution of
		  $\sum x_i/a_N$ converges to the stable distribution
		  with the characteristic exponent $\alpha$.
            \item When $\alpha=1$ in 
		  Eq.\
		  (\ref{necessary_condition_convergence_probability_distribution}), the probability distribution of
		  $\sum x_i/a_N-b_N$ converges to the stable distribution
		  with the characteristic exponent $\alpha=1$.
		  The centering constant $b_N$ diverges as
		  $N\rightarrow\infty$ \cite{footnote2}.  
	   \end{itemize}
 \end{theorem}

By using these theorems, we discuss sample-dependence of the dielectric
constant $\epsilon(\omega)\equiv\lim_{N\rightarrow\infty}\epsilon_N(\omega)$.
The discussion is divided into two parts: the case where the expectation 
value $ \langle 1/w_i \rangle$ is finite and the case where the
expectation value diverges.

When the expectation value $ \langle 1/w_i \rangle$ is finite, it is
shown from law of large numbers that 
\be
\epsilon(\omega)\equiv \lim_{N\rightarrow\infty}
\frac{1}{1+i \omega \sum_{i=1}^N (1/w_i)/N}=\frac{1}{1+i \omega 
\langle 1/w_i \rangle}.
\label{infty_toy}
\ee
Thus, there is no sample-dependence, i.e. the dielectric constant is 
self-averaging in the limit $N\rightarrow\infty$.
It is obvious that the Cole-Cole plot is the Debye 
semi-circle, which corresponds to the Cole-Cole plot of the dielectric 
constant for each sample. 

When the expectation value $ \langle1/w_i \rangle$ diverges (strong
disorder), we discuss below the two cases of the value of $\alpha$ in
Eq.\ (\ref{necessary_condition_convergence_probability_distribution}):
$0<\alpha<1$ and $\alpha=1$. 
In order to see frequency dependence of $\epsilon(\omega)$, the
macroscopic limit ($N\rightarrow\infty$) should be taken so that
$\omega\sum 1/w_i/N$ is finite. Otherwise, $\epsilon(\omega)$ becomes a
constant $0$ or $1$. 
\begin{itemize}
\item When $0<\alpha<1$, from the extended central limit theorem,
      $\tilde{T}\equiv \lim_{N\rightarrow\infty}\sum 1/w_i/a_N$ is
      a random variable with a limiting probability distribution 
      and hence it is finite with probability one. Hence, the limit
      $N\rightarrow\infty$ should be taken so that $\omega a_N/N$
      is finite;
      \bea
      \epsilon(\omega) & \equiv & \lim_{{N\rightarrow\infty}\atop{\;
      \tilde{\omega}=\omega a_N/N}} \frac{1}{1+i\omega \sum
      (1/w_i)/N} \nonumber \\
      & = & \lim_{{N\rightarrow\infty}\atop{\;
      \tilde{\omega} = \omega a_N/N}}\frac{1}{1+i\omega
      a_N/N \sum (1/w_i)/a_N} \nonumber \\
      & = & \frac{1}{1+i \tilde{\omega} 
      \tilde{T}}. 
      \eea
      It implies that the frequency-dependence is scaled with $\omega a_N/N$ 
      as shown in Fig.\ \ref{scaling}.
	\begin{figure}
\includegraphics[width=10cm,keepaspectratio]{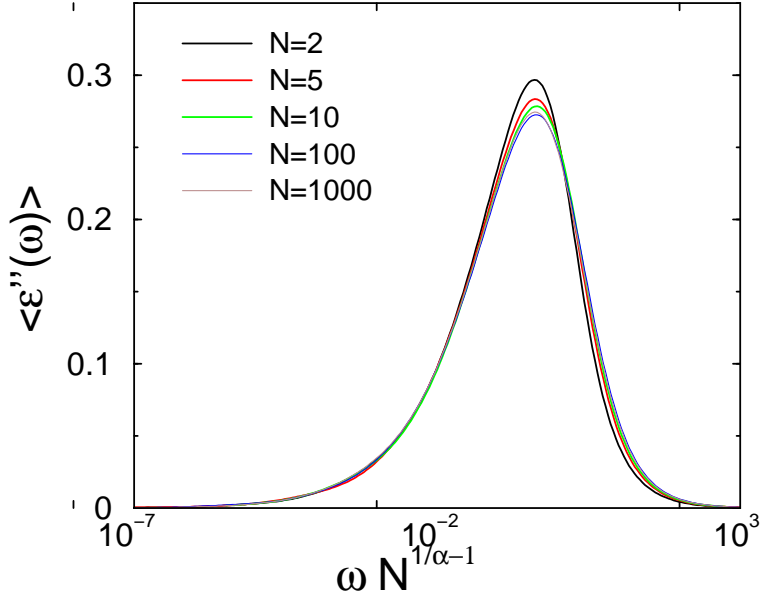}
\caption{The frequency-dependence of the imaginary part of 
the disorder-averaged dielectric 
constant when the probability distribution function of the jump rates 
is power-law $\rho(w_i)=\alpha w_i^{\alpha-1} \ (\alpha=1/2$). 
The two curves for sufficiently long chains $N=100, 1000$ lie on the 
same curve when plotted as a function of the scaled frequency $\omega 
N^{1/\alpha-1}$ ($\alpha=1/2$).}
\label{scaling}
\end{figure}

      Since $\tilde{T}$ is a random variable, $\epsilon(\omega)$ is also
      a random variable and hence $\epsilon(\omega)$ is non-self-averaging. In
      addition, since it is known that the probability density $\rho(x)$
      of a stable distribution with characteristic exponent $\alpha$
      behaves asymptotically as $\rho(x)\sim x^{-\alpha-1}$, 
      the probability density $\rho(\tilde{T})$ behaves
      as $\rho(\tilde{T})\sim {\tilde{T}}^{-\alpha-1}$. Hence, the expectation
      value $\langle \tilde{T} \rangle$ diverges. 
      Since it implies that the coefficient of the first order of low 
      $\tilde{\omega}$ expansion diverges, the disorder-averaged dielectric
      constant is a non-analytic function of $i \ \tilde{\omega}$ which
      must behave as $\langle \epsilon(\omega) \rangle -1 \sim 
      (i \ \tilde{\omega})^{\mu} \ (0<\mu<1)$. 
      Thus, from the same argument employed in the discussion for 
      the Cole-Cole plot of
      the admittance in Sec.\ \ref{comp_admitt_section} the 
      Cole-Cole plot of the 
      disorder-averaged dielectric constant appears inside the Debye
      semi-circle which corresponds to the Cole-Cole plot of the
      dielectric constant for each sample 
      (see Fig.\ \ref{cole-cole_de}).
\begin{figure}
\includegraphics[width=10cm,keepaspectratio]{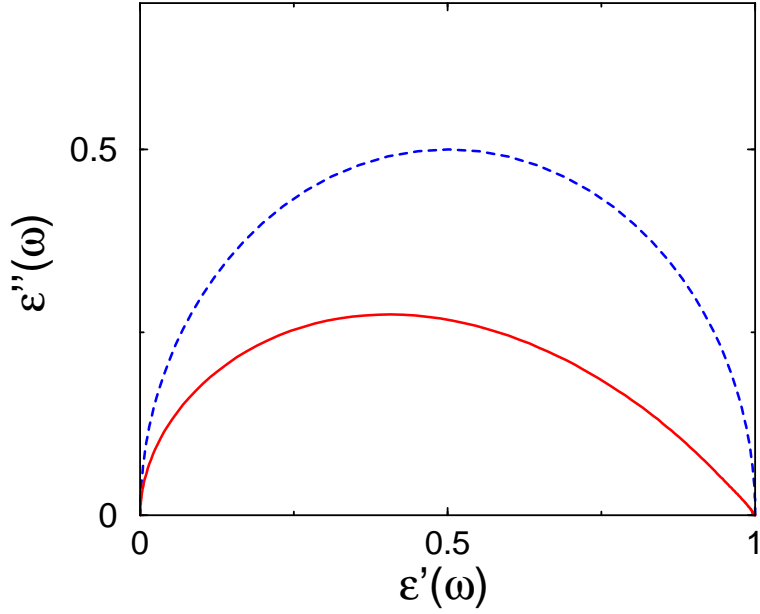}
\caption{The Cole-Cole plot of the disorder-averaged dielectric
 constant (solid line) when $w_i$ are random jump rates with the
 power-law distribution Eq.\ (\ref{jrdist}) with 
$\alpha=1/2$ and the Debye semi-circle (dashed line). 
It is seen that the Cole-Cole plot of 
the disorder-averaged dielectric constant is inside 
the Cole-Cole plot for each sample, i.e. the Debye semi-circle.}
\label{cole-cole_de}
\end{figure}

\item When $\alpha=1$, from the extended central limit theorem, 
      $\sum 1/(w_i)/a_N-b_N$ is a random variable with a limiting probability
      distribution and hence it is finite with probability one. 
      Furthermore, $b_N$ diverges as $N\rightarrow\infty$ as shown in
      appendix \ref{sec_centering_constant}. 
      Hence, the limit $N\rightarrow\infty$
      should be taken so that $\omega a_N b_N/N$ is finite;
      \bea
      \epsilon(\omega) & \equiv & \lim_{{N\rightarrow\infty}\atop{\;
      \tilde{\omega}=\omega a_N b_N/N}} \frac{1}{1+i\omega \sum (1/w_i)/N} 
      \nonumber \\
      & = & \lim_{{N\rightarrow\infty}\atop{\;\tilde{\omega}=\omega a_N b_N/N}}
      \frac{1}{1+i\omega a_N b_N/N\left[ \left( \sum (1/w_i)/a_N-b_N
      \right)/b_N +1 \right]} \nonumber \\
      & = & \frac{1}{1+i\tilde{\omega}}.
      \eea
      Hence, $\epsilon(\omega)$ is self-averaging.
\end{itemize}

Consequently, we conclude that {\it only when $0<\alpha<1$ 
the dielectric constant is non-self-averaging}. 

In order to test the foregoing discussion, 
we present an example of the above analysis where 
the random jump rates distributed by the power-law distribution 
Eq.\ (\ref{jrdist}) when $\alpha=1/2$.  
From the extended central limit theorem, 
it is known that the distribution function of $\tilde{T}$ is given by 
the stable distribution with $\alpha=1/2$. 
From Eq.\ (\ref{asymptotic_behavior_stable_distribution}), the
probability distribution $g(\tilde{T})$ of $\tilde{T}$ behaves as
$g(\tilde{T})\sim {\tilde{T}}^{-3/2}$. Hence, when $\tilde{\omega}\simeq
0$ the disorder-averaged dielectric constant behave as 
\be
\langle \epsilon(\omega) \rangle - 1 \sim \sqrt{i\tilde{\omega}}.
\ee 
On the other hand, the dielectric constant for each sample behaves as 
\be
\epsilon(\omega)-1 \sim i \tilde{\omega}.
\ee
Thus, the low-frequency behavior is non-self-averaging.
Absence of self-averaging is also seen in the Cole-Cole plot (Fig. 
\ref{cole-cole_de}).

\section{Higher dimensional disordered media}
\label{sec:h-d}

It is important to note that the system treated above 
is purely one-dimensional and it is still an open problem 
whether the admittance for higher dimensional media is self-averaging.

In this section, we investigate 
properties of the admittance for higher dimensional disordered media 
in the same manner for one-dimensional media. At first, the statistics 
of the MFPT for higher dimensional media is analyzed analytically. 
With the results, it is shown that the complex admittance is 
non-self-averaging. 
After that, the theoretical results are confirmed numerically. 

For its feasibility, 
we investigate a site-disordered model called the random trap model, which 
is defined as a random walk model where 
the jump rate $w_{i \ j}$ from site $j$ to site $i$ depends only on $j$.

\begin{figure}
\includegraphics[width=10cm,keepaspectratio]{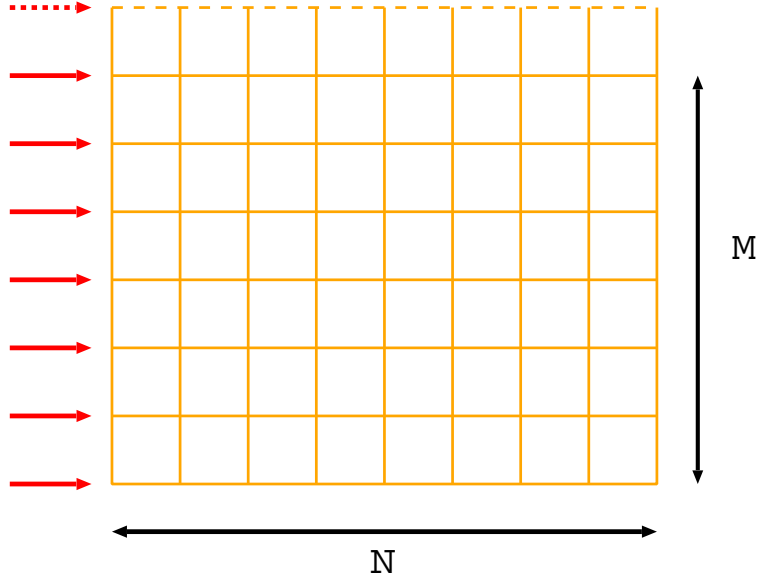}
\caption{The schematic picture of the two-dimensional medium 
with an oscillatory input current.}
\label{2-dimensional-lattice}
\end{figure}

We assume that the higher dimensional medium is a $N \times M$ 
square (or cubic) lattice. 
Here, $N$ is the linear dimension of the medium in direction 
of the current and a number of grid points on the cross section perpendicular 
to the current direction is denoted as $M$. 
The periodic boundary condition on the side ends is assumed, i.e., 
the medium is a hyper-cylinder, and the oscillatory input 
current is introduced uniformly on the left end of the medium.  
An example of two-dimension is illustrated in 
Fig.\ \ref{2-dimensional-lattice}.  
The probability distribution for the jump rates is assumed power-law
distribution Eq.\ (\ref{jrdist}) with the exponent $\alpha$.
Since we are interested in amorphous semiconductors, our analyses are limited 
to the case $0<\alpha<1$.

\subsection{The analytic expression for the MFPT of 
higher dimensional media}

At first, we introduce a general analytical expression for the MFPT in 
terms of the backward equation. After that, 
by using the expression, the results for the one-dimensional MFPT 
and the extended central limit theorem, the probability distribution of 
the higher-dimensional MFPT is investigated. 
Since the analyses are general, results are applicable both to 
two-dimensional and three-dimensional media. 

The backward equation, i.e., the time-development equation for a 
averaged quantity, is derived in the following way. 

Let $P_{\mathbf{j} \mathbf{i}}(t)$ be the probability that the 
random walker starting from the site $\mathbf{i}$ at time $0$ is 
found at the site $\mathbf{j}$ at time $t$. 
We assume that the master equation for the probability is written as 
\be
\dot{\mathbf{P_{\mathbf{i}}}}(t) = H \mathbf{P_{\mathbf{i}}}(t).
\ee
Here, $\mathbf{P_{\mathbf{i}}}$ is the vector whose $\mathbf{j}$-th 
component is $P_{\mathbf{j} \mathbf{i}}$ with fixing the starting site 
$\mathbf{i}$. 
Furthermore, $H$ is a linear operator given in terms of the jump rates. 

We consider the time-development equation of the average of 
the site-dependent quantity $f_{\mathbf{j}}$. The average $g_{\mathbf{i}}$ 
is defined by 
\be
g_{\mathbf{i}}(t) \equiv \sum_{\mathbf{j}} f_{\mathbf{j}} 
P_{\mathbf{j} \mathbf{i}}(t)
\ee
and hence the time-derivative is written as 
\be
\dot{g}_{\mathbf{i}}(t) = \sum_{\mathbf{j}} f_{\mathbf{j}} 
\dot{P}_{\mathbf{j} \mathbf{i}}(t) = \sum_{\mathbf{j} \mathbf{k}} 
f_{\mathbf{j}} H_{\mathbf{j} \mathbf{k}} 
P_{\mathbf{k} \mathbf{i}}(t).
\ee
Taken the limit $t\rightarrow 0$, 
\be
\dot{g}_{\mathbf{i}}(0) = \sum_{\mathbf{j} \mathbf{k}} 
f_{\mathbf{j}} H_{\mathbf{j} \mathbf{k}} \delta_{\mathbf{k} \mathbf{i}} = 
\sum_{\mathbf{j}} f_{\mathbf{j}} H_{\mathbf{j} \mathbf{i}} = 
\sum_{\mathbf{j}} \left( ^tH \right)_{\mathbf{i} \mathbf{j}}f_{\mathbf{j}}.
\ee
Let $\mathbf{g}(t)$ be the vector whose component is 
$g_{\mathbf{i}}(t)$, then 
\be
\dot{\mathbf{g}(0)} = ^tH \mathbf{f}.
\ee
By using the above result, $\mathbf{g}(t)$ is Maclaurin-expanded as 
\be
\mathbf{g}(t) = \sum_{n=0}^{\infty} \frac{1}{n!} \mathbf{g}^{(n)}(0) t^n = 
\sum_{n=0}^{\infty} \frac{1}{n!} (^tH)^n \mathbf{f} t^n.
\ee
Taken the time-derivative, 
\be
\dot{\mathbf{g}}(t) = \sum_{n=1}^{\infty}\frac{n}{n!} (^tH)^n \mathbf{f} 
t^{n-1} = ^tH \sum_{n=0}^{\infty} \frac{1}{n!}^n \mathbf{f} t^n = 
^tH \mathbf{g}(t).
\ee
The last equality gives the backward equation: 
\be
\dot{\mathbf{g}}(t) = ^tH \mathbf{g}(t).
\ee

As an example of the average $g_{\mathbf{i}}(t)$, we consider the probability 
that the random walker staring at the site $\mathbf{i}$ on the end 
illuminated with the laser light is found in the medium at the time 
$t ( >0 )$. The survival probability $n_{\mathbf{i}}(t)$ is defined as 
\be
n_{\mathbf{i}}(t) \equiv \sum_{\mathbf{j}} P_{\mathbf{j} \mathbf{i}}(t).
\ee 
Since the probability can be considered as a averaged quantity, 
$n_{\mathbf{i}}(t)$ obeys the backward equation:
\be
\dot{\mathbf{n}}(t) = ^tH \mathbf{n}(t),
\ee
where $\mathbf{n}(t)$ is the vector whose component is $n_i(t)$.

In terms of the survival probability, the first passage time distribution 
(FPTD) $F(t)$ is written as 
\be
F(t) = -\frac{d}{dt} \sum_{\mathbf{i}} n_{\mathbf{i}}(t)/M,
\ee
where $M \equiv \sum_{\mathbf{i}}1$ and the sum is taken over 
the grid points on the end of the medium. 
Hence, the MFPT $\bar{t}$ is also written as 
\be
\bar{t} \equiv \int^{\infty}_0 t F(t) dt = -\frac{1}{M} \sum_{\mathbf{i}} 
\int_0^{\infty} t \frac{d}{dt} n_{\mathbf{i}}(t) dt = 
\frac{1}{M} \sum_{\mathbf{i}} \int_0^{\infty} n_{\mathbf{i}}(t) dt = 
\frac{1}{M} \tilde{n}_{\mathbf{i}}(0).
\label{new-expression-MFPT}
\ee
Here, $\tilde{n}_{\mathbf{i}}(0)$ is the limit $s\rightarrow 0$ of the 
Laplace transform $\tilde{n}_{\mathbf{i}}(s)$. 
The Laplace-transform of the backward equation is written as 
\be
s \tilde{\mathbf{n}}(s)-\mathbf{n}(t=0) = ^tH \tilde{\mathbf{n}}(s).
\ee
Taken the limit $s\rightarrow 0$, 
\be
^tH \tilde{\mathbf{n}}(0) = -\mathbf{n}(t=0),
\label{laplace-transform-backward-eq}
\ee
where $\mathbf{n}(t=0)= ^t(1,1,\cdots)$.
By solving Eq.\ (\ref{laplace-transform-backward-eq}), the MFPT is obtained 
as Eq.\ (\ref{new-expression-MFPT}).

In the rest of this section, we concentrate on the random trap model. 
For the $d$-dimensional random trap model, the analytical expression of the 
MFPT can be easily obtained by Eq.\ (\ref{laplace-transform-backward-eq}) 
in the following way. 

Each components in the same \emph{column} of the matrix $H$ are written only 
in terms of the jump rates of the same site. It means that each components 
in the same \emph{line} of the transposed matrix $^tH$ are written only with 
the jump rates of the same site. Hence, divided the each lines of 
Eq.\ (\ref{laplace-transform-backward-eq}) by the jump rates, we obtain 
\bea
^t H_0 \tilde{\mathbf{n}}(0) = - \left( \begin{array}{c} \vdots \\ 
1/w_{\mathbf{k}} \\ \vdots \end{array} \right).
\label{rt-laplace-transform-backward-eq}
\eea
Here, $^tH_0$ is the transpose of the time-development operator when the 
jump rates of the all sites are equal to unity. From 
Eq.\ (\ref{rt-laplace-transform-backward-eq}), $\tilde{\mathbf{n}}(0)$ is 
a homogeneous linear function of the inverse of the jump rates.

We denote the grid points in $d$-dimensional medium with two indexes, 
$n$ and $\mathbf{m}$. One index $n$ is the label of the cross sections 
normal to the current direction and the other index $\mathbf{m}$ is the 
$d-1$ dimensional vector to denote a grid point on the cross section. 
Let the cross section at the left end, 
which is illuminated with laser light, be numbered $0$ and 
the cross section at the right end be numbered $N-1$. 
Each cross sections have $M$ grid points. 

Since the boundary condition at the side ends is periodic, each grid points 
in the same cross section are identical geometrically and each jump rates 
$w_{n \mathbf{m}}$ in the same cross section $n$ have the same contribution 
to the MFPT. Hence, the results obtained above means that the MFPT for 
the $d$-dimensional medium, $\bar{t}_d$ is written as 
\bea
\bar{t}_d & = & \frac{1}{M}\sum_{\mathbf{m}} \sum_{n=0}^{N-1}\frac{a_n}{w_{n 
\mathbf{m}}} \nonumber \\
&=& \frac{1}{M}\sum_{\mathbf{m}}\left( \frac{a_0}{w_{0 \mathbf{m}}}
+\frac{a_1}{w_{1 \mathbf{m}}}+\cdots+\frac{a_{N-1}}{w_{N-1 \mathbf{m}}}\right),
\label{mfpt-dd-rt-a}
\eea
where $\{a_n\}$ are constants independent of configuration of disorder. 
If all the jump rates in a cross section are equal, i.e., 
$w_{n \mathbf{m}}=w_{n \mathbf{m}'}\equiv w_n$, the $d$-dimensional MFPT 
$\bar{t}_d$ is equal to the one-dimensional MFPT $\bar{t}_1$, 
which is given as 
\be
\bar{t}_1 = \frac{N}{w_0}+\frac{N-1}{w_1}+\cdots +\frac{1}{w_{N-1}}.
\label{mfpt-1d-rt}
\ee
Comparison of Eq.\ (\ref{mfpt-dd-rt-a}) to Eq.\ (\ref{mfpt-1d-rt}) shows that 
\be
a_n = N-n.
\ee
Consequently, the MFPT for $d$-dimensional medium is given as 
\bea
\bar{t}_d & = & \frac{1}{M}\sum_{\mathbf{m}} \sum_{n=0}^{N-1}\frac{N-n}{w_{n 
\mathbf{m}}} \nonumber \\
&=& \frac{1}{M}\sum_{\mathbf{m}}\left( \frac{N}{w_{0 \mathbf{m}}}
+\frac{N-1}{w_{1 \mathbf{m}}}+\cdots+\frac{1}{w_{N-1 \mathbf{m}}}\right).
\label{mfpt-dd-rt}
\eea
With respect to the MFPT, the $d$-dimensional medium is a bundle of 
one-dimensional chain and hence the MFPT is rewritten as 
\be
\bar{t}_d = \frac{1}{M} \sum_{\mathbf{m}} t^{(\mathbf{m})}_1, 
\ee
where $t^{(\mathbf{m})}_1$ is the MFPT for a one-dimensional medium defined as 
\be
t^{(\mathbf{m})}_1 \equiv \sum_{n=0}^{N-1} \frac{N-n}{w_{n \mathbf{m}}}.
\ee

\subsection{The statistics of the MFPT for higher dimensional media}
\label{sec:pdf-mfpt-dd}

In order to consider the probability distribution of $\bar{t}_d$, 
the following postulate, which is expected to be true from the results of 
analyses for the one-dimensional case, is introduced: 
Let $\{ u_n \}$ be a set of independent and identically distributed random 
variables. The tail of the probability density is assumed to be 
\be
\rho_u(u_n)\sim {u_n}^{-\alpha-1},
\ee
where $0<\alpha <1$.
Then, we postulate that the random variable $v$ defined as 
\be
v \equiv \frac{1}{N^{1/\alpha+1}} \sum_{n=0}^{N-1} (N-n) u_n
\label{definition-v}
\ee
has a normalized probability distribution function and the tail obeys the 
power law.

By making use of the postulate, we analyze the statistics of $\bar{t}_d$ 
in two different ways and determine the asymptotic form of the probability 
distribution of $\bar{t}_d$. 

At first, we take the way where the sum in terms of $n$ is done first. 
From our postulate, the random variable $\tilde{t}_{\mathbf{m}}$ defined as 
\be
\tilde{t}_{\mathbf{m}} \equiv \frac{1}{N^{1/\alpha+1}} t^{(\mathbf{m})}_1
\ee
has the normalized probability distribution function and it obeys the power 
law:
\be
\rho_{\tilde{t}}(\tilde{t}_{\mathbf{m}}) \sim 
{\tilde{t}_{\mathbf{m}}}^{-\alpha'-1},
\ee
where $\alpha'$ is a positive constant. Since $\bar{t}_d = M^{-1} 
N^{1/\alpha+1} \sum_{\mathbf{m}}\tilde{t}_{\mathbf{m}}$, 
from the extended central limit theorem, the random variable $x$ defined as 
\be
x \equiv \frac{1}{M^{1/\alpha'-1} N^{1/\alpha+1}} \bar{t}_d
\label{normalization1}
\ee
has a normalized probability distribution function and its tail 
asymptotically behaves as 
\be
\rho_x(x) \sim x^{-\alpha'-1}.
\ee

As the other way, we evaluate the sum in terms of $\mathbf{m}$ first. 
The extended central limit theorem shows that 
the random variable $y_n$ defined as 
\be
y_n \equiv \frac{1}{M^{1/\alpha}} \sum_{\mathbf{m}} \frac{1}{w_{n \mathbf{m}}}
\ee
has a normalized probability distribution function whose asymptotic form is 
\be
\rho_y(y_n) \sim {y_n}^{-\alpha-1}.
\ee
Furthermore, since $\bar{t}_d = M^{1/\alpha-1}\sum_{n=0}^{N-1}(N-n) y_n$, 
our postulate means that the random variable $z$ defined as 
\be
z \equiv \frac{1}{M^{1/\alpha-1} N^{1/\alpha+1}} \bar{t}_d
\label{normalization2}
\ee
has a probability distribution function whose tail behaves as 
\be
\rho_z(z) \sim z^{-\alpha-1}.
\ee

Here, we compare the two ways of normalization Eqs.\ (\ref{normalization1}), 
(\ref{normalization2}). Since the two ways of normalization to obtain the 
normalized distribution function should be identical, 
\be
\alpha' = \alpha.
\label{equivalence-alpha}
\ee
Summarized our results, the MFPT $\bar{t}_d$ is normalized as 
\be
\bar{t}_d \equiv M^{1/\alpha-1} N^{1/\alpha+1} z
\label{normalized-MFPT-dd}
\ee
and the probability distribution function has the power-law tail as 
\be
\rho_z(z) \sim z^{-\alpha-1}.
\label{tail-normalized-MFPT-dd}
\ee
Furthermore, from Eq.\ (\ref{equivalence-alpha}), the tail of the probability 
distribution function of $v$ defined by Eq.\ (\ref{definition-v}) behaves as 
\be
\rho_v(v) \sim v^{-\alpha-1}.
\ee

\begin{figure}
\includegraphics[width=10cm,keepaspectratio]{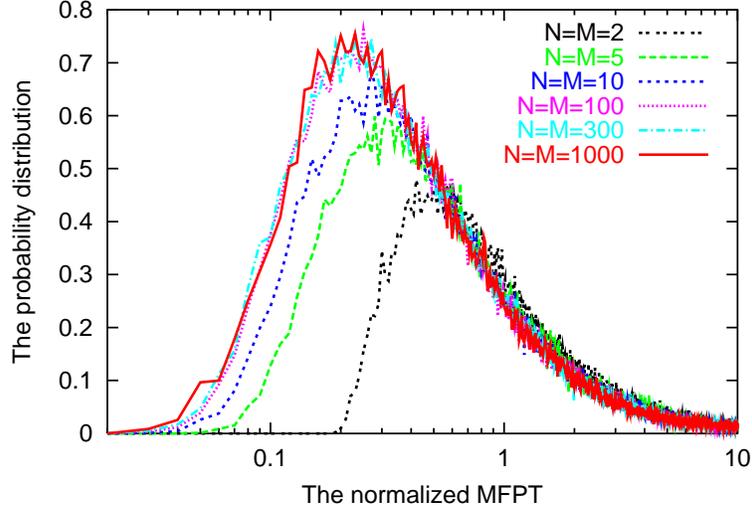}
\caption{The probability distribution function of the normalized MFPT 
	 $\bar{t}_d N^{-1/\alpha-1} M^{-1/\alpha+1}$ for
	 the random trap model where the probability distribution function of
	 the random jump rates is power-law Eq.\ (\ref{jrdist}) 
	with exponent $\alpha=1/2$. 
	The distribution is computed from 100000 samples 
	of the length and width $N, M=2, 5, 10, 100, 300, 1000$. 
	It is clearly seen that the probability 
	distribution functions lie on a same curve when the length of lattice
	 $N$ is sufficiently large.}
\label{pdf-mfpt-2d}
\end{figure}

\begin{figure}
\includegraphics[width=10cm,keepaspectratio]{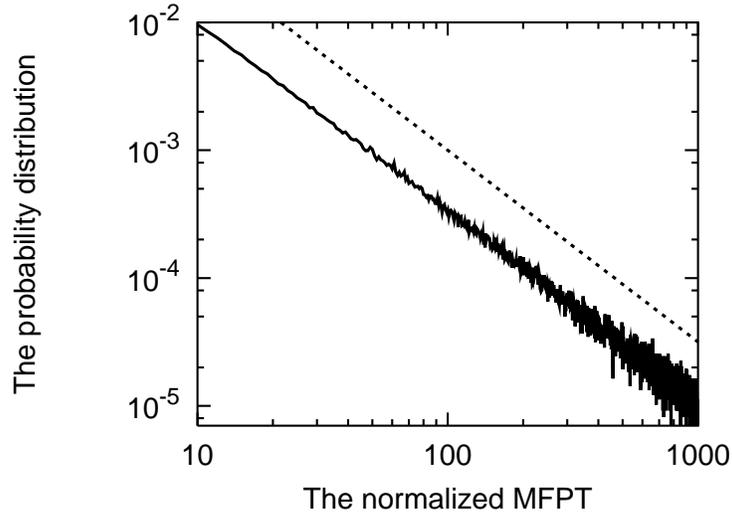}
\caption{The asymptotic behavior of the probability distribution
 function of the normalized MFPT computed from 1000000 samples. 
 The solid line denotes the
 distribution function when $N=M=100$ and $\alpha=1/2$ 
 and the dashed line represents the
 function proportional to $z^{-1.5}$. One sees clearly that the
 distribution function behaves asymptotically as $\rho_z(z) \sim z^{-\alpha-1} 
(-\alpha-1 \simeq 1.5 < 2).$}
\label{pdf-mfpt-100x100-tail}
\end{figure}

Since the results are based on the postulate, we confirm numerically them. 
Figure\ \ref{pdf-mfpt-2d} shows that the probability distribution function of 
$z$ converges to a unique distribution when $N$ and $M$ are increased. 
The power-law tail of the distribution is shown in 
Fig.\ \ref{pdf-mfpt-100x100-tail}. 
It is clearly seen that the exponent is equal to $-1.5$ when $\alpha=0.5$.

\subsection{Non-self-averaging complex admittances for 
higher dimensional media}

By using the results obtained in the previous subsection, 
Eqs.\ (\ref{normalized-MFPT-dd}), (\ref{tail-normalized-MFPT-dd}), we show in 
the same way as for the one-dimensional case that the complex admittance for 
higher dimensional media is also non-self-averaging when $0<\alpha<1$. 

The normalization of the MFPT, Eq.\ (\ref{normalized-MFPT-dd}), means that 
the complex admittance for an infinitely large higher dimensional medium is 
given as 
\be
\lim_{N,M\rightarrow \infty} \chi_{N,M}(\omega) = 1-i z \tilde{\omega}+\ldots 
\ee
in terms of the scaled frequency $\tilde{\omega}$ defined as 
\be
\tilde{\omega} \equiv \omega M^{1/\alpha-1} N^{1/\alpha+1}.
\ee
Here, the infinite volume limit $N,M\rightarrow \infty$ is taken with fixing 
$\tilde{\omega}$ at a finite value. 
Since the normalized MFPT $z$ is a random number, the complex admittance is 
dependent on each samples even for the infinitely large medium. 
That is, \emph{the complex admittance is non-self-averaging.} 

\begin{figure}
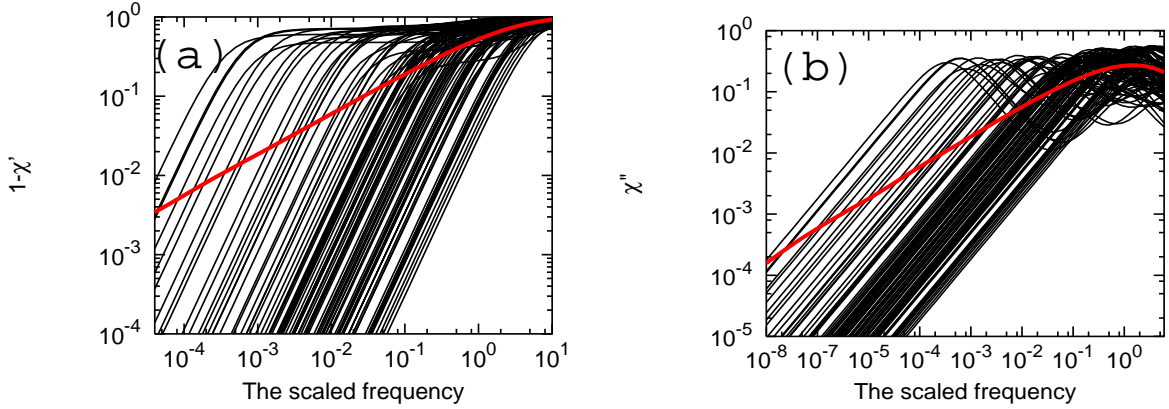

\includegraphics[width=8cm,keepaspectratio]{Fig17a.eps}
\includegraphics[width=8cm,keepaspectratio]{Fig17b.eps}
\caption{The frequency dependence of the real part (a) and 
the imaginary part (b) of the complex admittances for two-dimensional media 
when the exponent of the probability distribution of jump rates 
is $\alpha=1/2$ and the size of the media is $N=M=10$.
The thin solid lines represent the admittances for each of 100 samples.
The bold solid line represents the admittance averaged over 20000 samples.
Though the admittance of each sample is proportional to 
the scaled frequency $\tilde{\omega}$,
the averaged admittance is proportional to $\sqrt{\tilde{\omega}}$ 
($\mu \simeq 1/2$).}
\label{are-aim}
\end{figure}

Furthermore, from Eq.\ (\ref{tail-normalized-MFPT-dd}), it is shown that 
$\langle z \rangle = \infty$. It means non-analyticity of the complex 
admittance about $\tilde{\omega}=0$ and hence 
\be
\langle \lim_{N,M\rightarrow \infty} \chi_{N,M}(\omega) \rangle-1 \sim 
(i \tilde{\omega})^{\mu},
\ee
where $\mu$ is a constant such that $0<\mu<1$ (see 
Fig.\ \ref{are-aim}). 

\begin{figure}
\includegraphics[width=6cm,keepaspectratio]{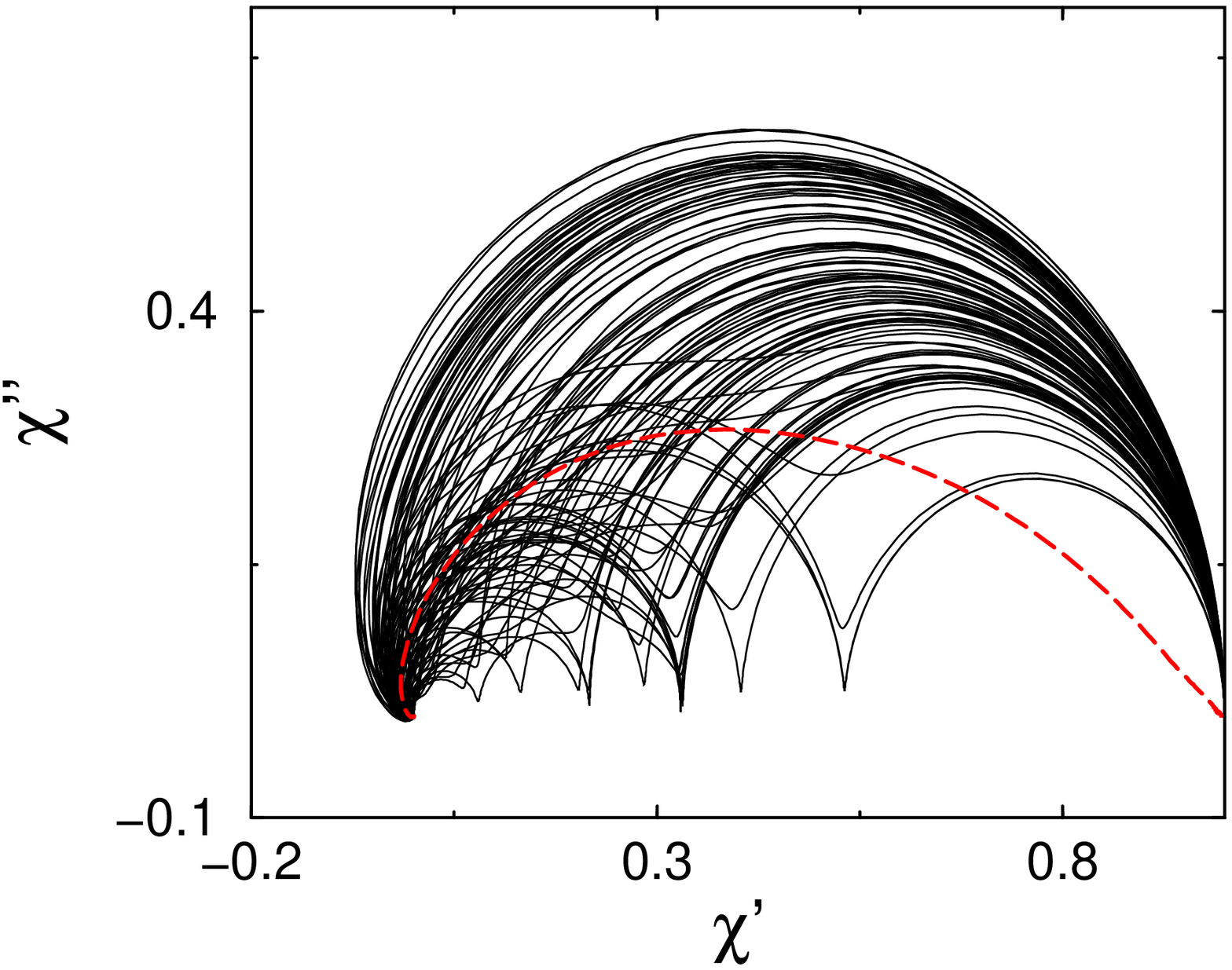} 
\includegraphics[width=6cm,keepaspectratio]{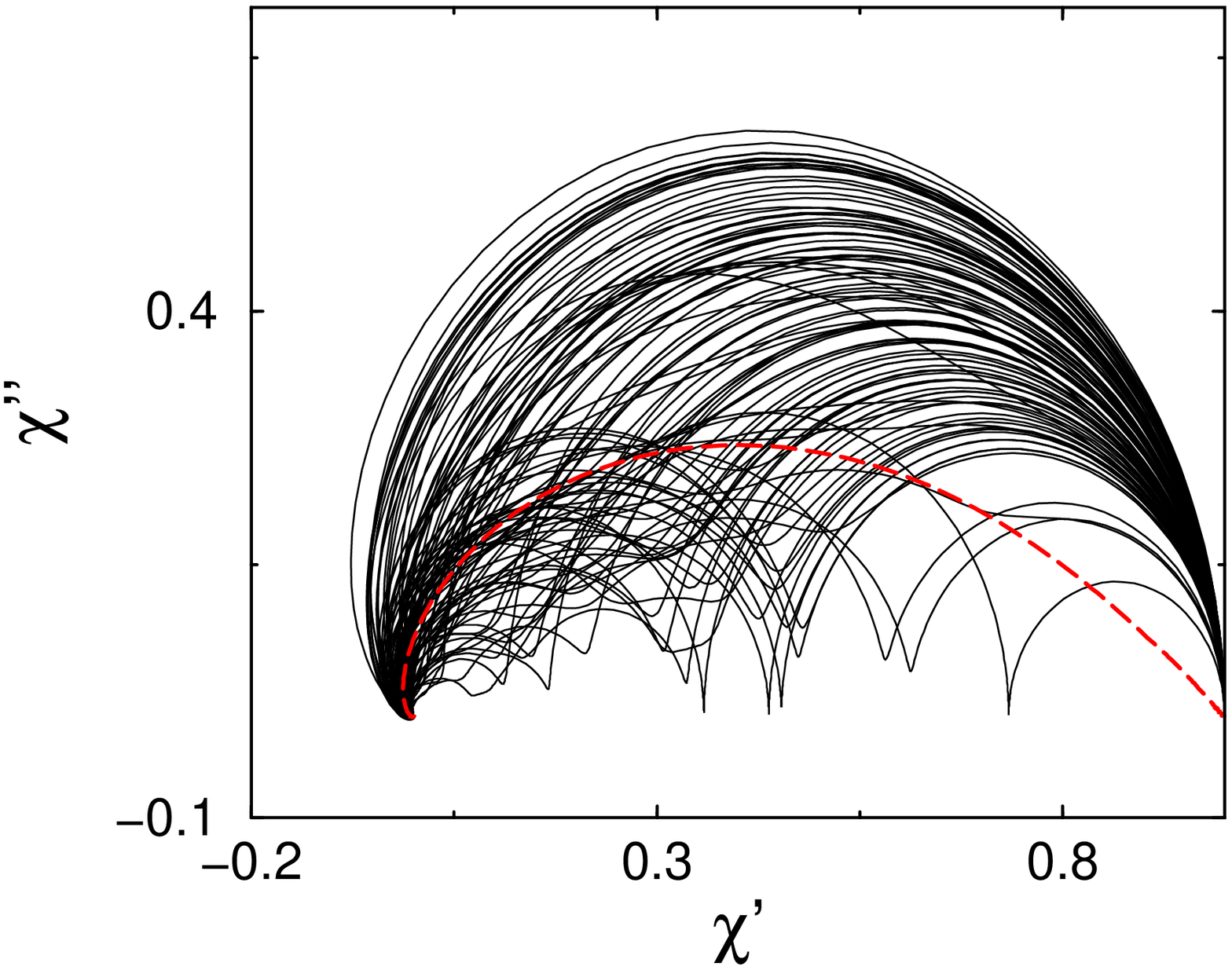} 
\includegraphics[width=6cm,keepaspectratio]{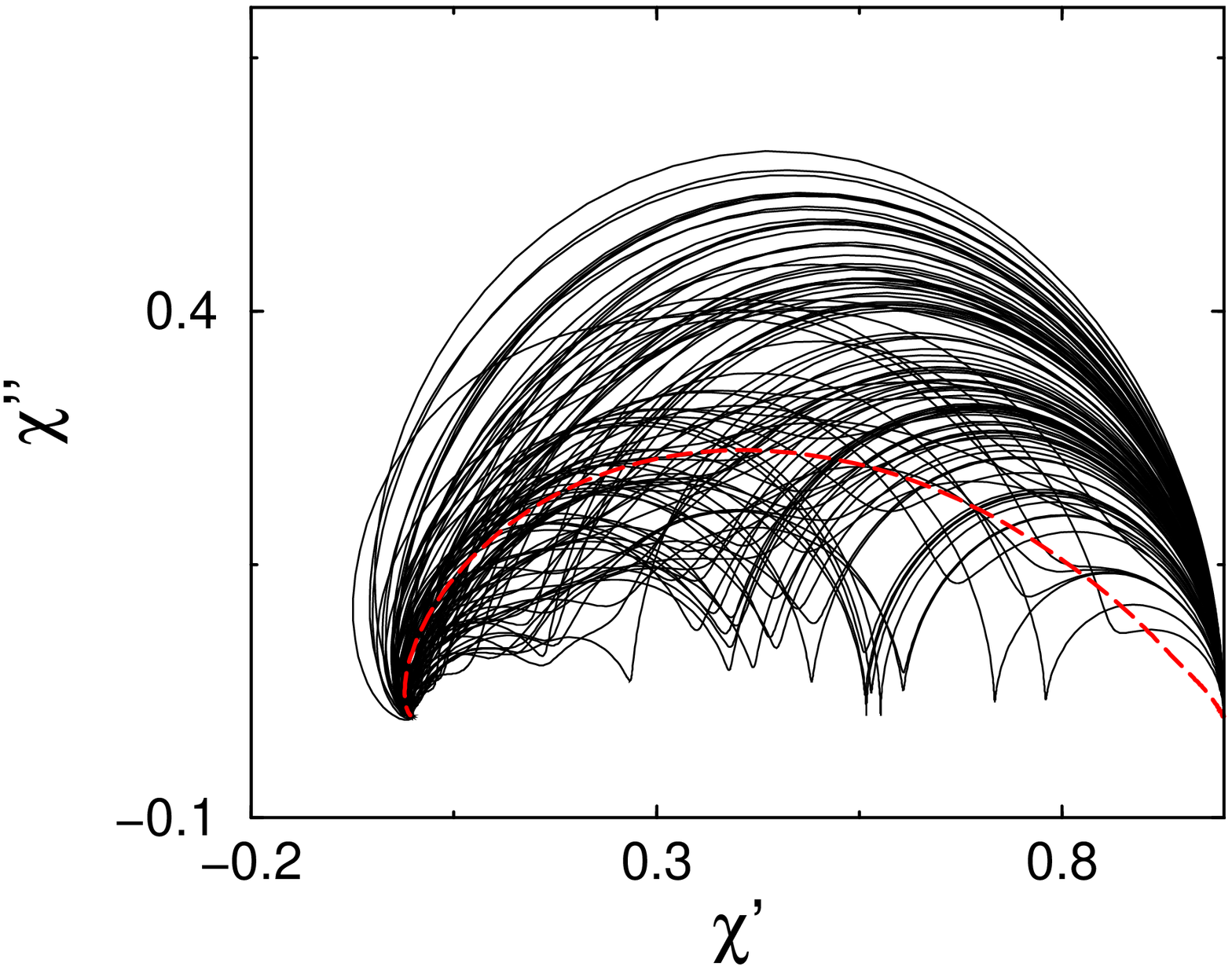}
\caption{(a) The Cole-Cole plot for $4 \times 4$ disordered square
 lattice. 
 The probability distribution for the jump rates is power-law
 distribution Eq.\ (\ref{jrdist}) when $\alpha=1/2$.
 The solid lines denote admittances for 100 samples and 
 the dashed line denotes the admittance disorder-averaged over 5000 samples. 
 (b) The Cole-Cole plot for $10 \times 10$ disordered square lattice. 
 (c) The Cole-Cole plot for $30 \times 30$ disordered square
 lattice. 
 These numerical results show that the sample-dependence of the 
 admittance seems to exist even when the size of the lattice is large and 
 hence it means that the admittance is non-self-averaging.
 }
\label{yxy}
\end{figure}

The results obtained above are all concerned with the low frequency behavior. 
We investigate numerically the non-self-averaging characteristics in the 
all frequency range with the Cole-Cole plot.  
Figure\ \ref{yxy} shows the Cole-Cole plots of 
the admittances for one hundred realizations of 
$4\times 4, 10\times 10, 30\times 30$ square lattices.
These numerical results show that the sample-dependence of the 
admittance seems to exist even when both the length of the lattice and
the width are large and 
hence it implies that {\it the admittance is non-self-averaging even in 
higher dimensional cases.} 

\section{Discussion and conclusions}
\label{sec:conclusion}

In conclusion, we presented a generic stochastic dynamical model in 
one-dimensional medium of the BPM and showed that the admittance 
has important information about transport in disordered media, e.g. the FPTD.
Thus, the BPM will be a powerful technique to investigate 
the dynamical process in random media.
We also showed that the admittance is non-self-averaging when the jump 
rates are random variables with the power-law distribution as $\rho(w_i)\sim
{w_i}^{\alpha-1} \ (0<\alpha\leq 1)$. 
It is important to note that this power-law distribution is realized in
amorphous semiconductors. 

On the other hand, when $1 \leq \alpha$, the low frequency behavior of 
$O(\omega)$ has no sample dependence as shown in Eqs.\ (\ref{adm.alpha=1}), 
(\ref{adm.alpha.g.1}).
However, with respect to the terms of $O(\omega^2)$, the 
second moment of the inverse of the jump rate 
$\langle 1/w^2 \rangle$, which diverges when $1 \leq \alpha \leq 2$, 
is concerned.
It implies that some anomalous statistical properties may be observed. 
Properties for the full frequency range were analyzed with the 
Cole-Cole plots. In the marginal case when $\alpha=1$, the non-self-averaging 
property appears as shown in Fig.\ \ref{Cole-Cole.1.0}. 
However, Fig.\ \ref{Cole-Cole2} shows that such effects from terms of 
$O(\omega^2)$ are small when $1<\alpha$.

When the complex admittance is non-self-averaging, 
the behavior of the averaged admittance over infinite number of samples 
does not coincide with that of 
the admittance observed in experiments. 
In order to see the non-self-averaging behavior described above 
one needs to work in one-dimensional system or the case where 
the translational invariance is violated in only one direction. 

Furthermore, in order to investigate 
possibility that the non-self-averaging properties of the admittance are 
observed in real experiments on higher dimensional media, 
we performed theoretical and numerical analyses of the admittance on 
higher dimensional disordered media. 
The analyses on anomalous statistics of the MFPT show that the 
non-self-averaging characteristics are observed in the low-frequency 
behavior of the complex admittance of higher-dimensional media. 
The numerical results confirm the conclusion that the admittance is 
non-self-averaging. Hence, it suggests possibility 
that the non-self-averaging properties of the admittance of the BPM may 
be observed in the real experiments on higher dimensional media.

On the other hand, non-self-averaging behavior, i.e. strong sample-dependence 
may be observed in other experiments. For example, 
it is known that anomalous system-size dependence of the mobility, 
$\mu \sim L^{1-1/\alpha}$, is observed in the  
time-of-flight experiment of amorphous semiconductors \cite{Pfister78}.
The anomalous system-size dependence is due to the fact that 
the smallest value of random jump rates depends on the length of a chain 
\cite{Kehr98}. Since the smallest value 
dominates the mobility, the mobility is also a random variable.
Thus, the anomalous system-size dependence of the mobility implies that the 
mobility is not self-averaging.

\acknowledgments

I wish to acknowledge valuable discussions with 
Akira Yoshimori, Hiizu Nakanishi, Kazuo Kitahara, Osamu Narikiyo and 
Miki Matsuo. One of us (M. K.) is thankful to 
Institute f\"{u}r Festk\"{o}rperforschung, 
Forschungszentrum J\"{u}lich for its hospitality, 
where part of the present work was done.
This work was supported in part by the Grant-in-Aid for Scientific 
Research of Ministry of Education, Culture, Sports, Science and Technology.

\appendix

\newtheorem{definition}{Definition}[section]

\section*{Law of large numbers and the extended central limit theorem}
\label{appendix}

In this appendix, we summarize the notions of probability theory, 
partly rewritten for our convenience, used to 
analyze the complex admittance and the ``dielectric constant'' in Secs.\
\ref{comp_admitt_section} and \ref{toy_section}.
After that, we show that $b_N$ Eq.\ (\ref{b_N}) 
diverges as $N\rightarrow\infty$.

\subsection{The extended central limit theorem}
\label{appendix:eclt}

Here, we summarize theorems on probability distribution of normalized sum of
random variables \cite{Feller66}.

The following 
law of large numbers tells that a mean converges to an expectation
value even when a variance does not exist.
In this section, $S_n$ denotes the sum $\sum_{i=1}^n X_i$.

\begin{theorem}[Strong law of large numbers]
Let the $X_k$ be independent random variables 
with a common distribution $F$. If they
 have an expectation $\mu$ then $S_n/n \rightarrow \mu$ with
 probability one. 
\end{theorem}

In order to describe the extended central limit theorem, some concepts
are defined.

\begin{definition}[Domain of attraction]
The distribution $F$ of the independent random variables $X_k$ belongs
 to the domain of attraction of a distribution $R$ if there exist
 the norming constant $a_n>0$ and the centering constant 
 $b_n$ such that the distribution of
 $a_n^{-1}S_n-b_n$ tends to $R$. 
\end{definition}

\begin{definition}[Slow variation]
A positive function $L$ defined on $(0,\infty)$ varies slowly at
 infinity if 
\be
\frac{L(s t)}{L(s)}\rightarrow 1
\label{def_slow_variation}
\ee
for each $t>0$ as $s\rightarrow\infty$.
\end{definition}
It implies that the leading term of $L(st)$ is $L(s)$ when $s$ is large.
Constant functions and $\log(x)$ are examples.

In addition, the function $F(x)$ is assumed to be related to the
probability density $\rho(x)$ as 
\be
F(x)\equiv\int_{-\infty}^x dy \rho(y).
\label{def_F}
\ee

Using definitions presented above, 
the extended central limit theorem is stated as follows:

\begin{theorem}[The extended central limit theorem]
 In order that $F$ belongs to some domain of attraction it is necessary
 that $F$ is of the form 
 \be
 1-F(x)+F(-x)\sim x^{-\alpha} L(x)
 \label{condition_convergence}
 \ee
 for some $0<\alpha\leq 2$ as $x\rightarrow\infty$.
 \begin{itemize}
  \item If $\alpha=2$ then $F$ belongs to the domain of attraction of the
	normal distribution.
  \item If $\alpha <2$ and 
	\be
	\frac{1-F(x)}{1-F(x)+F(-x)}\rightarrow p, \ 
	\frac{F(-x)}{1-F(x)+F(-x)}\rightarrow q
	\label{condition_q_p}
	\ee
	as $x\rightarrow\infty$ then $F$ belongs to the domain of attraction of
	the stable distribution with the characteristic exponent
	$\alpha$. If Eq.\ (\ref{condition_q_p}) fails then $F$ belongs to
	no domain of attraction.
\end{itemize}
\end{theorem}
Here, Eq.\ (\ref{condition_q_p}) only means that the limits exist. 
If positive random variables are considered, $p=1$ and $q=0$ since $F(-x)=0$. 
The limits $p$ and $q$ give parameters of a stable distribution. 
In addition, the centering parameter $b_n$ is given as followings;
\begin{itemize}
\item If $\alpha>1$ then $b_n$ is given by the expectation value.
\item If $\alpha=1$ then $b_n$ is given by 
      \be
      b_n = n \int_{-\infty}^{\infty}dx \sin\left(\frac{x}{a_n}\right)\rho(x).
      \label{centering_constant}
      \ee
\item If $\alpha<1$ then $b_n=0$. 
\end{itemize}
The norming constant $a_n$ is chosen so that 
\be
\frac{n}{{a_n}^{\alpha}}L(a_n)\rightarrow C,
\label{norming_constant}
\ee
where $C>0$ is a constant.

This theorem implies that {\it when the disorder is strong and the
probability distribution function of the normalized sum $S_n/a_n-b_n$
converges 
in the limit $n\rightarrow\infty$ the limiting probability distribution 
is the stable distribution with the characteristic exponent
$0<\alpha\leq 1$ given by Eq.\ (\ref{condition_convergence}).}
The inequality $0<\alpha\leq 1$ is due to divergence of the expectation
value (the strong disorder).

For our convenience, we rewrite the necessary condition Eq.\
(\ref{condition_convergence}) for the
convergence of the probability distribution of $S_n/a_n-b_n$ in terms of
the probability density $\rho(x)$.
Since $\rho(x)=F'(x)\equiv dF(x)/dx$, the asymptotic behavior of the
probability density is given by 
\be
\rho(x)+\rho(-x) \sim x^{-\alpha-1}L(x)[1-\frac{1}{\alpha} \frac{x
L'(x)}{L(x)}].
\ee
The second term on the right hand side is evaluated from Eq.\
(\ref{def_slow_variation}). By differentiation with respect to $t$ of Eq.\
(\ref{def_slow_variation}) and setting $t=1$, it is shown that 
\be
\lim_{s\rightarrow\infty}\frac{L'(s)s}{L(s)}=0.
\ee 
It implies that 
\be
\rho(x)+\rho(-x) \sim x^{-\alpha-1}L(x)
\label{asymptotic_rho}
\ee
as $x\rightarrow\infty$.

For the asymptotic behavior of the stable distribution, the following
theorem is known.
\begin{theorem}[Asymptotic behavior of a stable distribution]
 If $G$ is a stable distribution obtained as the limiting distribution
 of the extended central limit theorem, then as $x\rightarrow\infty$ 
 \be
 x^{\alpha} [1-G(x)] \rightarrow C p \frac{2-\alpha}{\alpha}, \  
 x^{\alpha} G(-x) \rightarrow C q \frac{2-\alpha}{\alpha},
 \ee
 where the parameters $C, p, q$ are given by Eqs.\
 (\ref{condition_q_p}) and (\ref{norming_constant}).
\end{theorem}
It implies that the asymptotic behavior of the probability density
$g(x)=G'(x)$ is given by 
\be
g(x) \simeq C p (2-\alpha) x^{-\alpha-1}, \ 
g(-x) \simeq C q (2-\alpha) x^{-\alpha-1}.
\label{asymptotic_behavior_stable_distribution}
\ee
Hence, a stable distribution with the characteristic exponent $\alpha$
has absolute moments of all orders $<\alpha$ 
and absolute moments of all orders $\geq\alpha$ do not exist.

Next, we present the relation holding when $0<\alpha<1$.
It tells about the contribution of the maximal term to the sum.
Let $X_1,X_2,\ldots$ be independent random variables with the common
distribution $F$ satisfying Eq.\ (\ref{condition_convergence}) with
$0<\alpha<1$, i.e., belonging to the domain of attraction of the stable
distribution with the characteristic exponent $\alpha$. Put $M_n =
\max(X_1,\ldots, X_n)$. Then, 
\be
\left\langle \frac{S_n}{M_n}\right\rangle\rightarrow \frac{1}{1-\alpha}
\label{max_strong_disorder}
\ee
as $n\rightarrow\infty$.
It implies that the maximal term is of the same order as the sum with 
probability one when $0<\alpha<1$.

\subsection{The centering constant $b_n$ when $\alpha=1$}
\label{sec_centering_constant}

Here, we show that the centering constant $b_n$ diverges for
the positive random variables $x$ satisfying Eq.\
(\ref{condition_convergence}) with $\alpha=1$;
\be
1-F(x)\sim x^{-1} L(x).
\label{condition_convergence_alpha_1}
\ee

Let $x_0>0$ denote the lower limit of the 
distribution and the probability density is given by $\rho(x)=F'(x)$. 
From Eq.\ (\ref{norming_constant}), the norming constant $a_n$ is
chosen so that 
\be
\frac{n}{a_n}L(a_n)\rightarrow C,
\label{norming_constant_alpha_1}
\ee
where $C>0$ is a constant.
The asymptotic behavior of the probability density $\rho(x)$ is derived 
from Eq.\ (\ref{asymptotic_rho});
\be
\rho(x) \sim x^{-2} L(x).
\label{asymptotic_rho2}
\ee

We evaluate $b_n$ by dividing the region of integration of Eq.\
(\ref{centering_constant}); 
\be
b_n=n\int^{x_n}_{x_0}dx \sin\left(\frac{x}{a_n}\right) \rho(x)+
n\int^{\infty}_{x_n}dx \sin\left(\frac{x}{a_n}\right) \rho(x)
\label{b_n_2_division}
\ee
where $x_n$ is a constant satisfying 
\be
x_n \ll a_n, \ \ \lim_{n\rightarrow\infty}x_n=\infty.
\ee
Since $x_n \ll a_n$, the first term on the right hand side of Eq.\
(\ref{b_n_2_division}) is given by 
\be
\mbox{the first term}\simeq \frac{n}{a_n}\int^{x_n}_{x_0}dx x \rho(x) > 0.
\ee
From Eq.\ (\ref{asymptotic_rho2}), the leading term of 
the second term is evaluated as 
\bea
\mbox{the second term} & \sim & n\int_{x_n}^{\infty}dx
\sin\left(\frac{x}{a_n}\right)x^{-2}L(x) \nonumber \\
& = & \frac{n}{a_n}\int^{\infty}_{x_n/a_n}dx\sin\left(x\right) x^{-2}
L(a_n x),
\eea
where a numerical constant is ignored.
Since $L$ is a function slowly varying, the leading term is
given by 
\bea
\mbox{the second term} & \sim &
\frac{n}{a_n}L(a_n)\int^{\infty}_{x_n/a_n}dx
\sin\left(x\right)x^{-2} \nonumber \\
& \sim & \frac{n}{a_n}L(a_n)\log(a_n).
\eea
From Eq.\ (\ref{norming_constant_alpha_1}), it is shown that 
\be
\mbox{the second term}\sim \log(a_n).
\ee
Consequently, since the first term is positive and the second term
becomes positive infinity in the limit $n\rightarrow\infty$, 
the centering constant $b_n$
diverges in the limit $n\rightarrow\infty$.


\begin{thebibliography}{99}
\bibitem{footnote1}This power-law probability distribution for the jump 
rates corresponds to the dwell time distribution 
$\Psi(t) \sim t^{-\alpha-1}$ used in
the study of dispersive transport where $\alpha$ is proportional to 
temperature \cite{Pfister78}, 
and it has been shown that the power-law
distribution is common for activation processes with random activation
energy \cite{Odagaki95}.
\bibitem{footnote2}It is shown in appendix \ref{sec_centering_constant}.
\end{thebibliography}
\end{document}